\documentclass[fleqn,usenatbib]{mnras}

\usepackage[T1]{fontenc}
\usepackage{ae,aecompl}

\usepackage{graphicx}	
\usepackage{amsmath}	
\usepackage{amssymb}	

\usepackage{newtxtext,newtxmath}

\title[V339 Delphini during the 2013 outburst]{Temporal variability and obscuration effects in the X-ray emission of classical nova V339 Delphini (Nova Delphini 2013)}

\author[Pei et al.]{
Songpeng Pei,$^{1}$\thanks{E-mail: songpengpei@outlook.com}
Nataly Ospina,$^{2}$
Xiaowan Zhang,$^{1}$
Qiang Li,$^{3, 4}$
Ziwei Ou,$^{5}$
\newauthor
Taozhi Yang,$^{6}$
and Yongzhi Cai$^{7, 8, 9}$
\\
$^{1}$School of Physics and Electrical Engineering, Liupanshui Normal University, Liupanshui, Guizhou, 553004, China\\
$^{2}$Department of Theoretical Physics, University Autonoma Madrid, 28049 Madrid, Spain\\
$^{3}$Qiannan Normal University for Nationalities, Duyun 558000, China\\
$^{4}$Qiannan Key Laboratory of Radio Astronomy, Guizhou Province, Duyun 558000, China\\
$^{5}$Tsung-Dao Lee Institute, Shanghai Jiao Tong University, Shanghai 201210, China\\
$^{6}$Ministry of Education Key Laboratory for Nonequilibrium Synthesis and Modulation of Condensed Matter, School of Physics, Xi'an Jiaotong University,\\ 710049 Xi'an, China\\
$^{7}$Yunnan Observatories, Chinese Academy of Sciences, Kunming 650216, China\\
$^{8}$International Centre of Supernovae, Yunnan Key Laboratory, Kunming 650216, China\\
$^{9}$Key Laboratory for the Structure and Evolution of Celestial Objects, Chinese Academy of Sciences, Kunming 650216, China\\
}

\date{Accepted 00.00.2024. Received 00.00.2024; in original form 00.00.2024}

\pubyear{2015}

\begin{document}
\label{firstpage}
\pagerange{\pageref{firstpage}--\pageref{lastpage}}
\maketitle

\begin{abstract}
In this study, we present a detailed analysis of public archival soft X-ray data on the classical nova V339 Delphini (Nova Del 2013) during its outburst, obtained using the {\it Chandra} High-Resolution Camera Spectrometer (HRC-S) and Low Energy Transmission Grating (LETG), as well as {\it XMM-Newton} in 2013. The observations, spanning from day 85.2 to day 112.0 after the optical maximum, capture the nova during its luminous supersoft X-ray source (SSS) phase. The spectra reveal numerous absorption features with blue-shifted velocities ranging from $\sim$ 724 to $\sim$ 1474 km s$^{-1}$, with the majority of lines blue-shifted by approximately 1200 km s$^{-1}$. We confirm the presence of a short-period modulation of the X-ray flux with a period of approximately 54 seconds, as well as the drift of this period, which was detected on days 97.0 and 112.0 during the outburst with both {\it XMM-Newton} and {\it Chandra}. This period modulation is transient in nature, with significant variations in amplitude and pulse profile over timescales of a few thousand seconds, likely due to temporary obscuration events that affect the emission from the central hot source. The pulse profiles exhibit substantial deviations from a pure sinusoidal shape, which may be related to the period drift. Additionally, the modulation amplitude shows a possible anti-correlation with the count rates on day 97.0, likely also caused by temporary obscuration events influencing the central source's emission.
\end{abstract}



\section{Introduction} \label{sec:intro}
Classical nova V339 Delphini (V339 Del; Nova Delphini 2013; PNV J20233073+2046041) was discovered by Koichi Itagaki at a magnitude of V = 6.8 on August 14.584 Universal Time (UT), 2013 (Julian Day (JD) 2456519.084) \citep{2013CBET.3628....1N}. The nova explosion began on August 13.9 UT, 2013 (JD 2456518.4) \citep{2015ApJ...812..132G}. Approximately 1.666 days after its discovery, the nova reached its optical maximum brightness at magnitude $V\sim 4.3$ on August 16.25 UT, 2013 (JD 2456520.75) \citep{2013ATel.5304....1M}, making it a rare nova observable with the naked eye in the Northern sky. The estimated decline times from maximum brightness for a decay of 2 and 3 magnitudes in the V band were $t_{\rm 2,V}$ = 10 days and $t_{\rm 3,V}$ = 18 days, respectively, indicating that V339 Delphini (hereafter V339 Del) was a fast nova \citep{2014CoSka..43..330C}.

\citet{2013IAUC.9258....2D} identified the progenitor of V339 Del as the blue star USNO-B1.0 1107-0509795, with magnitudes of $B \sim 17.20-17.4$ and $R \sim 17.45-17.74$. The total outburst amplitude in V was measured at 12.6 magnitudes \citep{2013ATel.5297....1M}. \citet{2013IBVS.6087....1M} also located the progenitor on Asiago 1979-1983 photographic plates, noting that it was marginally detected by the APASS all-sky survey\footnote{http://www.aavso.org/apass} in April 2012. They also found that the brightness and colour of the progenitor remained stable for an extended period before the 2013 eruption, with the total amplitude of variation in the B-band being 0.9 magnitudes over approximately four years of observations, from 1979-04-23 to 1983-04-09, indicating that the progenitor was dominated by the emission originating from an accretion disc. Additionally, \citet{2014A&A...563A.129D} found that the progenitor did not exhibit significant variability on a timescale of approximately 0.5 hours in the 1.2 years leading up to the 2013 eruption.

Explosive element lithium (Li) production was detected in V339 Del, making it the first nova observed to synthesize Li, supporting theoretical predictions that Li can be produced in novae \citep{2015Natur.518..381T}. \citet{2015A&A...581A.134D} analyzed high-resolution optical spectra and identified the primary star as a carbon-oxygen (CO) WD. They also found that the structure of the nova ejecta is non-spherical and inhomogeneous. Low-resolution spectroscopy confirmed that V339 Del is a typical Fe II nova \citep{2015BaltA..24..109B}. \citet{2014CoSka..43..330C} estimated the mass of the central WD to be $M_{\rm WD} = 1.04~\pm 0.02M_{\rm \odot}$ based on $M_{\rm B,max}$, while \citet{2024ApJ...965...49H} determined a significantly larger mass of $1.25~\pm 0.05M_{\rm \odot}$ using the maximum magnitude versus rate of decline (MMRD) relationship. The orbital period of the binary system was initially estimated to be between 3.15 and 6.43 hours \citep{2014CoSka..43..330C, 2015ASPC..496..237C} and later measured as $0.162941 \pm 0.000060$ days ($3.910584 \pm 0.001440$ hours) \citep{2022MNRAS.517.3640S}.

V339 Del was observed in $\gamma$-rays \citep[][]{2014Sci...345..554A,2015A&A...582A..67A}, X-rays \citep[][]{2013ATel.5429....1P,2013ATel.5470....1P, 2013ATel.5505....1O, 2013ATel.5573....1B, 2013ATel.5593....1N, 2013ATel.5626....1N, 2014ATel.5967....1P}, ultraviolet \citep[][]{2013ATel.5409....1S,2013ATel.5624....1S, 2014ATel.6088....1S, 2016A&A...590A.123S}, optical \citep[][]{2013ATel.5378....1S,2013ATel.5546....1S,2013IBVS.6080....1M,2013ATel.5624....1S, 2014CoSka..43..330C,2014A&A...569A.112S, 2015BaltA..24..109B, 2015Natur.518..381T, 2015NewA...40...28M, 2016AstL...42...10T, 2016A&A...590A.123S, 2017Ap.....60...19S, 2019ApJ...872..120K,2019RMxAA..55..141J}, near-infrared \citep[][]{2013ATel.5404....1B}, infrared \citep[][]{2014AstL...40..120T,2015ApJ...812..132G}, radio and millimeter wavelengths \citep[][]{2013ATel.5382....1C,2013ATel.5428....1A}.

\begin{figure*}
\centering
\includegraphics[width=17.0cm]{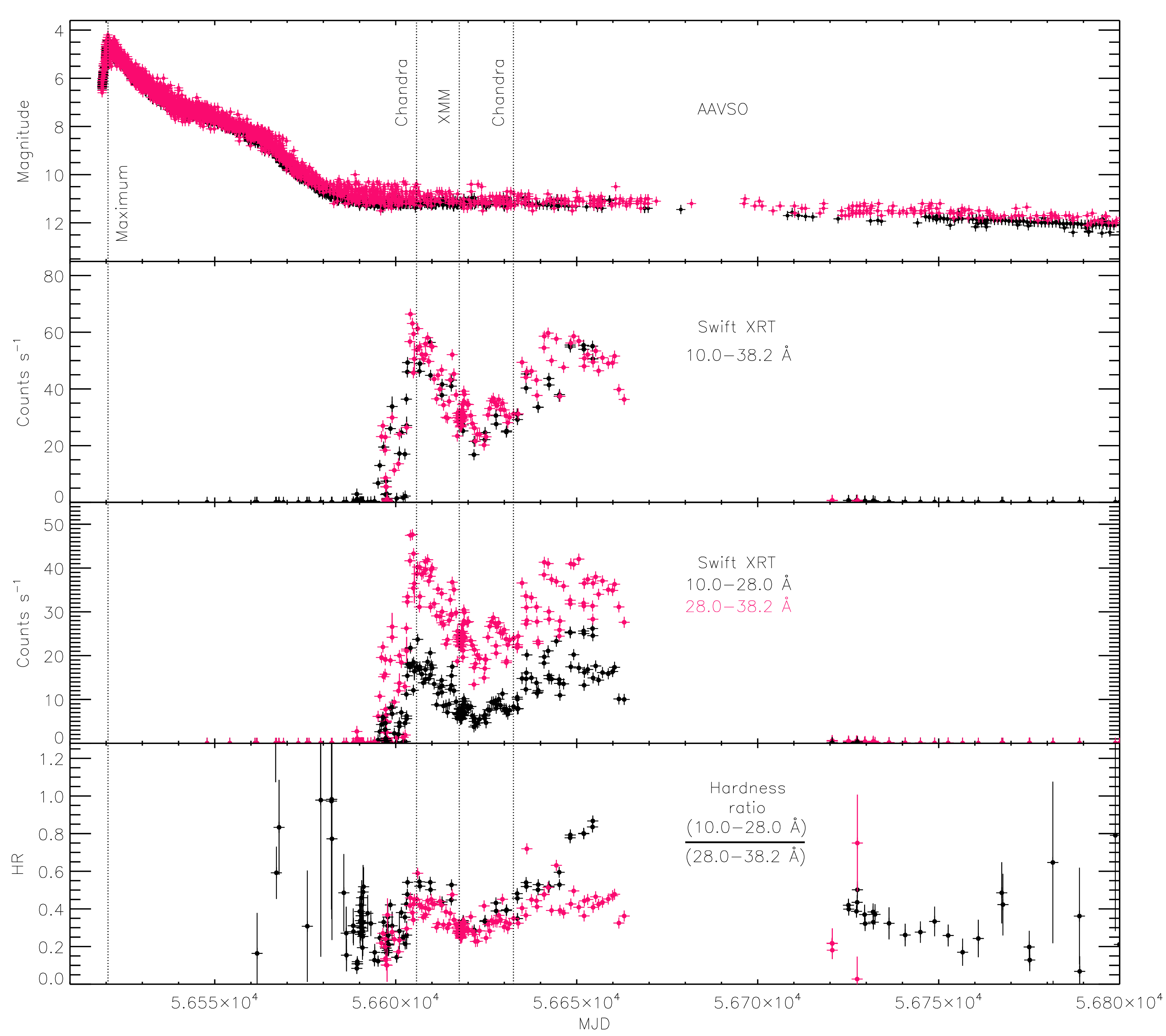}
\caption{The top panel displays the American Association of Variable Star Observers (AAVSO) visual light curve of V339 Del (red) alongside its AAVSO V-band light curve (black). The first vertical line in all panels marks the peak of the optical outburst (MJD 56520.25), the third vertical line marks the times of the {\it XMM-Newton} observations, and the remaining vertical lines indicate the times of the {\it Chandra} observations. The lower panel shows the {\it Swift} XRT light curve in the 10.0 $-$ 38.2 \AA\ band (0.32 $-$ 1.24 keV) in Photon Counting (PC) mode (black) and Windowed Timing (WT) mode (red). The remaining panels display the {\it Swift} XRT light curves in the 10.0 $-$ 28.0 \AA (black) and 28.0 $-$ 38.2 \AA (red) bands, along with the X-ray hardness ratio: (10.0 -- 28.0 \AA)/(28.0 -- 38.2 \AA) from data obtained in PC (black) and WT (red) modes.}\label{fig:sum}
\end{figure*}

Fig.~\ref{fig:sum} shows the optical light curve of the nova from a $\sim$ 0.95 days after its explosion to day 280 after the optical maximum. The interstellar reddening to V339 Del has been determined to be $E_{\rm B-V}$ = 0.182 \citep{2013ATel.5297....1M} and $0.184 \pm 0.035$ \citep{2014CoSka..43..330C}. The distance to the nova, derived from the observed parallax in {\sl Gaia} data release 2 (DR2), is estimated to be 2.13$^{+2.25}_{-0.40}$\,kpc \citep[][]{2018MNRAS.481.3033S} and 2.53$^{+2.23}_{-1.11}$\,kpc \footnote{from the GAIA database using ARI's Gaia Services at \url{http://gaia.ari.uni-heidelberg.de/tap.html.}}. The distance derived from the observed parallax in {\sl Gaia} early data release 3 (eDR3) is 2.06$^{+1.22}_{-0.75}$\,kpc\footnote{http://dc.g-vo.org/tableinfo/gedr3dist.main} \citep[][]{2021AJ....161..147B} and 1587$^{+1338}_{-299}$\,pc \citep{2022MNRAS.517.6150S}. Similarly, \citet{2024ApJ...965...49H} obtained a distance of $2.1 \pm 0.2$\,kpc, adopting $E_{\rm B-V}$ = 0.18.

\citet{2014A&A...569A.112S} estimated the effective temperature (T$_{\rm eff}$) of the white dwarf's pseudophotosphere to be between 6000 and 12000 K by modeling the optical/near-IR spectral energy distribution (SED) during the fireball stage (August 14.8 $-$ 19.9, 2013; days 0.9 $-$ 6.0 after the nova explosion began). Comparing the SED of V339 Del in the 1.25 $-$ 5 $\upmu$m range on two observation dates (August 15.94 and 16.86, 2013 UT; days 2 and 3 after the explosion) to the SEDs of normal supergiants (B5I and A0I), \citet{2014AstL...40..120T} estimated the nova's temperature to be $\sim$ 13600 and $\sim$ 9400 K, respectively. During the SSS phase, \citet{2013ATel.5593....1N} modeled the first high-resolution X-ray spectrum of the nova, obtained by the LETG/HRC-S instrument on board the {\it Chandra} observatory on November 09.75, 2013 (day 85.2 after the optical maximum), using a simple absorbed blackbody model. They found a photospheric temperature of the WD to be 27 eV ($\sim$ 310000\,K). From the spectrum obtained with European Space Agency’s X-ray Multi-Mirror Mission ({\it XMM-Newton}) on November 21, 2013 (day 97.0 after the optical maximum), \citet{2013ATel.5626....1N} estimated the temperature of this SSS to be around 30 eV ($\sim$ 350000\,K) based on the Wien tail. \citet{2023ApJ...943...31M} estimated a rough T$_{\rm eff}$ of $\sim$ 600,000 K using a two-shell model to fit the high-resolution X-ray spectra of V339 Del, extracted from two {\it Chandra} observations on days 85.2 and 112.0 after the optical maximum.

In this work, we reanalyzed previously published data obtained with {\it Chandra} \citep{2023ApJ...943...31M} and {\it XMM-Newton} \citep{2015A&A...578A..39N}. The X-ray spectra of V339 Del are complex, with features in the high-resolution grating spectra that could be invaluable for constraining models. However, previous attempts at spectral modeling of these high-resolution X-ray spectra have been only partially successful, and the high-resolution X-ray spectrum obtained with {\it XMM-Newton} has not yet been thoroughly analyzed. Therefore, it is valuable to revisit the high-resolution X-ray spectra of V339 Del, incorporating timing analysis. In Sect. \ref{sec:intro}, we provide a brief introduction to V339 Del. The details of the observations and a summary of the data reduction process are described in Sect. \ref{sec:observation}. A detailed timing analysis is presented in Sect. \ref{sec:timing}, followed by a thorough analysis of the grating spectra in Sect. \ref{sec:spectral}. We discuss our findings in Sect. \ref{sec:discussion} and summarize our conclusions in Sect. \ref{sec:conclusions}.


\section{Observation and data reduction} \label{sec:observation}
V339 Del was observed twice in 2013 by the {\it Chandra} observatory using the HRC-S \citep{2002PASP..114....1W} and the LETG \citep{2000SPIE.4012....2W, 2008HEAD...10.0403D}. Additionally, it was observed once by the {\it XMM-Newton} observatory \citep{2001A&A...365L...1J, 2012OptEn..51a1009L} using the European Photon Imaging Camera-Metal Oxide Semiconductor (EPIC-MOS) \citep{2001A&A...365L..27T} and the Reflection Grating Spectrometers \citep[RGS;][]{1998sxmm.confE...2B, 2001A&A...365L...7D}. The details of these observations are summarized in Table~\ref{table:obs}. The principal investigator for the {\it Chandra} observations was Thomas John Nelson \citep{2013ATel.5593....1N}, while Gregory Schwarz led the {\it XMM-Newton} observation. The nominal exposure times for the three observations (15742, 0728200201, and 15743) were 50,180 s, 33,820 s, and 49,550 s, respectively. However, the dead-time corrected exposures (net exposure times) were slightly shorter, as shown in Table~\ref{table:obs}.
 
We extracted the {\it Chandra} zero-order light curve and the first-order spectrum using {\it Chandra} Interactive Analysis of Observations (CIAO) v4.12 \citep{2006SPIE.6270E..1VF} \footnote{https://cxc.cfa.harvard.edu/ciao/download/}, along with calibration files CALDB v4.9.0. All standard data reprocessing and reduction procedures from CIAO science threads were applied. The background-subtracted light curves from the {\it Chandra} HRC-S camera (zero-order) were extracted from event files using the `DMEXTRACT' tool. High-resolution spectra from the {\it Chandra} HRC-S+LETG were obtained using the `chandra$_{-}$repro' script, and the `combine$_{-}$grating$_{-}$spectra' script was used to combine the first-order grating redistribution matrix files (RMFs), ancillary response files (ARFs), and grating spectra.

We performed the {\it XMM-Newton} data reduction using the Science Analysis System (SAS) version 20.0.0 package and the Current Calibration Files (CCF). The spectra and light curves for RGS1, RGS2, and MOS were extracted using the updated calibration files and following the standard data reprocessing and reduction procedures outlined in the SAS Data Analysis Threads. The background-subtracted grating spectra from RGS1 and RGS2 were then combined and averaged using the SAS task RGSCOMBINE.

This classical nova was also monitored by the Neil Gehrels {\it Swift} Observatory \citep[hereafter, {\it Swift};][]{2004ApJ...611.1005G}. The {\it Swift} X-ray Telescope \citep[XRT;][]{2005SSRv..120..165B} light curves were obtained using the online {\it Swift} XRT data products generator \footnote{https://www.swift.ac.uk/user\_objects/} \citep{2007A&A...469..379E, 2009MNRAS.397.1177E}. High-resolution X-ray spectra were analyzed and fitted using the command-driven, interactive X-ray spectral fitting package XSPEC (version 12.12.1) \citep{1996ASPC..101...17A, 2003HEAD....7.2210D}.


\begin{table*}
\begin{minipage}{180mm}
\caption{{\it Chandra} and {\it XMM-Newton} observations of V339 Del analysed in this work, and measured count rates for the X-ray detectors. The reported errors are at the 90\% confidence level.
}
\label{table:obs}
\begin{center}
\begin{tabular}{lllllllll}\hline\hline \noalign{\smallskip}

 Instrument & Date$^a$ & ObsID & Day$^b$ &    Exp. time$^c$ & C.R. (LETG)  & C.R. (HRC-S or & F$_{\rm x}$  $\times 10^{-9}$ \\
            &  (UTC)     &           &           & (ks)         & (counts s$^{-1}$) &  RGS) (counts s$^{-1}$) &   (erg s$^{-1}$ cm$^{-2}$) \\

\hline \noalign{\smallskip}
 {\it Chandra} HRC-S+LETG & 2013 Nov 09     & 15742 & 85.2 &  45.869  & 68.21$\pm$0.06     &  47.18$^d$    & 3.44$^e$  \\
 {\it XMM-Newton} RGS & 2013 Nov 21     & 0728200201 & 97.0 &  32.590  & $...$                 &  62.88$^f$    &    1.38$^g$   \\
 {\it Chandra} HRC-S+LETG & 2013 Dec 06     & 15743 & 112.0 &  48.914  & 48.25$\pm$0.05   &   33.00$^d$   & 2.42$^e$  \\
\hline \noalign{\smallskip}
\end{tabular}
\end{center}
Notes:\hspace{0.1cm} $^a $: Start date of the observation. $^b $: Time in days after the optical peak on August 16.25 UT, 2013 (Modified Julian Date (MJD) 56520.25). $^c $: Observation exposure time, corrected for dead time. $^d$: Measured count rates for the HRC-S. $^e $: X-ray flux measured between 18 and 124 \AA \ by integrating the flux from the LETG. $^f $: Measured count rates for the RGS. $^g $: X-ray flux measured between 18 and 38.2 \AA \ by integrating the flux from the RGS. \\
\end{minipage}
\end{table*}
\begin{figure*}
\centering
\includegraphics[width=17.0cm]{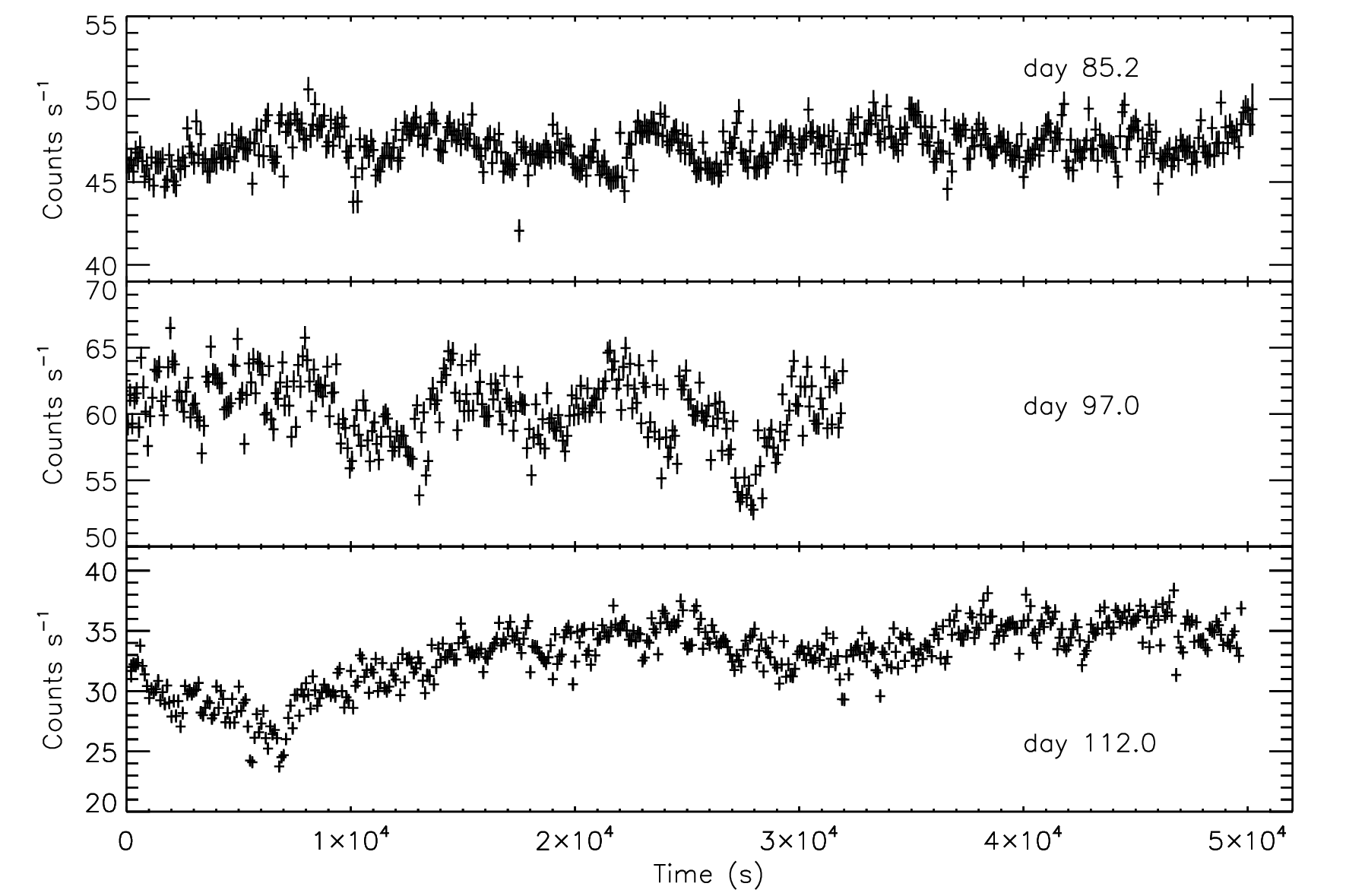}
\caption{The background-subtracted zero-order light curves of V339 Del, measured with the {\it Chandra} HRC-S camera and {\it XMM-Newton} RGS on days 85.2, 97.0, and 112.0, were binned every 100 seconds. Note that different y-axis scales have been used for these three exposures. The RGS wavelength range is 6 $-$ 38.2 \AA \, corresponding to the 0.324 $-$ 2.066 keV energy range, with the source signal exceeding the background only between 155 and 324 eV. The HRC-S camera is calibrated for the 0.06 $-$ 10 keV range; however, based on the LETG spectrum, the source signal was above the background only between 155 and 560 eV.}\label{fig:lc}
\end{figure*}
\begin{figure*}
\begin{center}
\includegraphics[width=88mm]{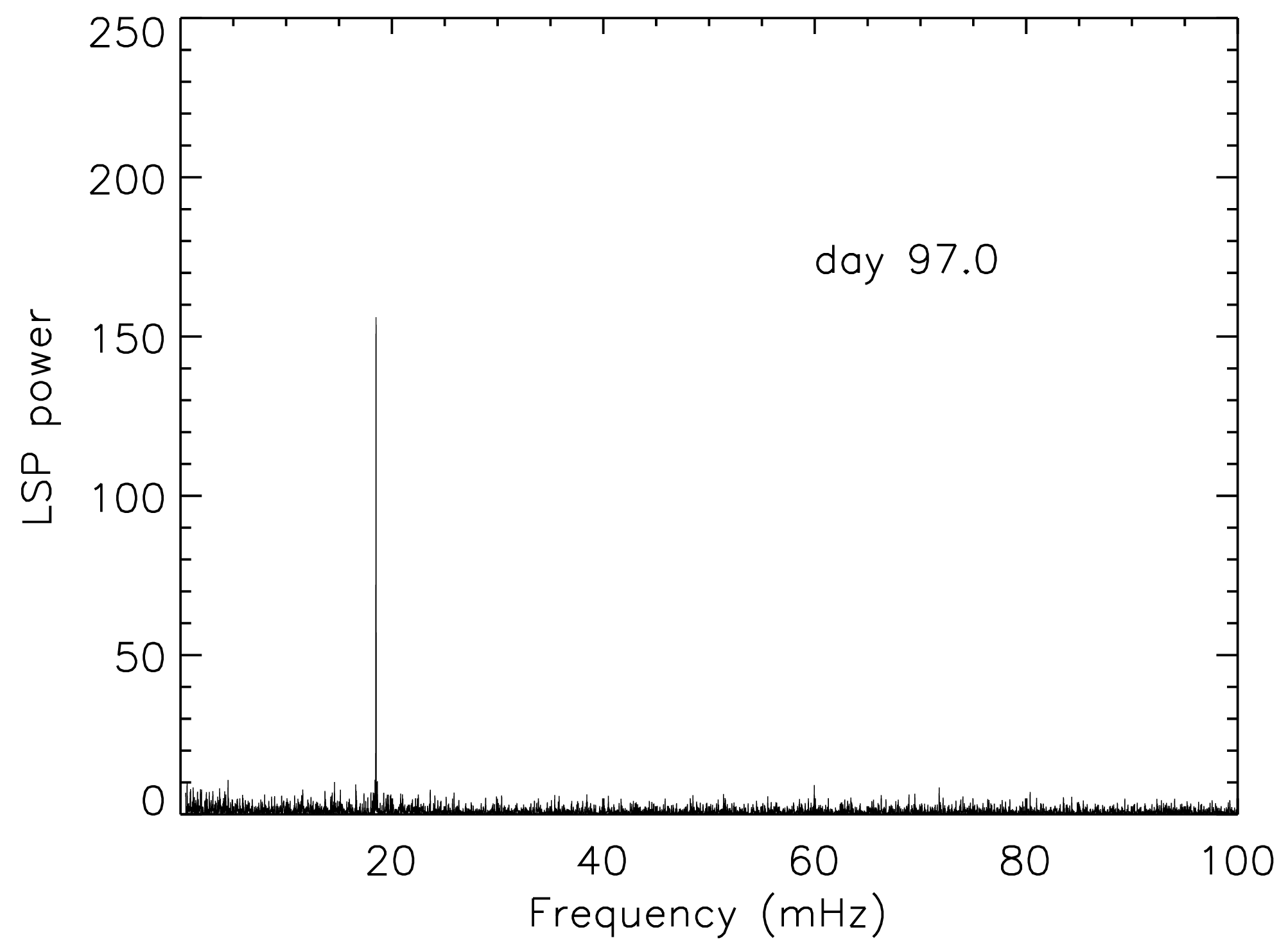}
\hspace{-0.62em}
\includegraphics[width=88mm]{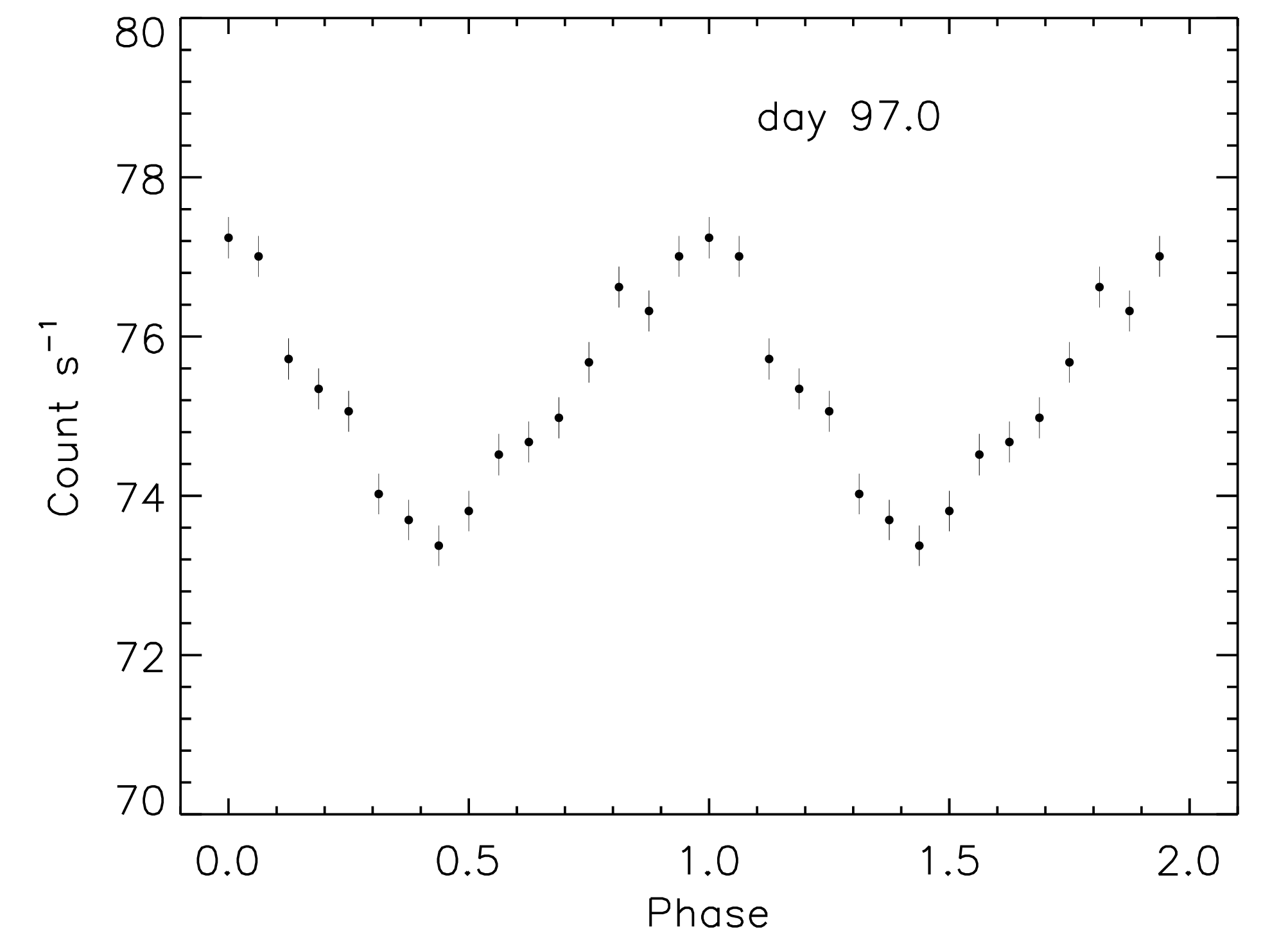}

\caption{Left: Lomb-Scargle periodogram of the MOS 2 light curve from the {\it XMM-Newton} observation 0728200201 of V339 Del. Right: The light curve folded with the detected period of 54.06 seconds (corresponding to $\simeq$ 18.50 mHz). The epoch of the first data point serves as the zero point.}
\label{fig: lsp and pfold lc xmm}
\end{center}
\end{figure*}
\begin{figure*}
\begin{center}
\includegraphics[width=88mm]{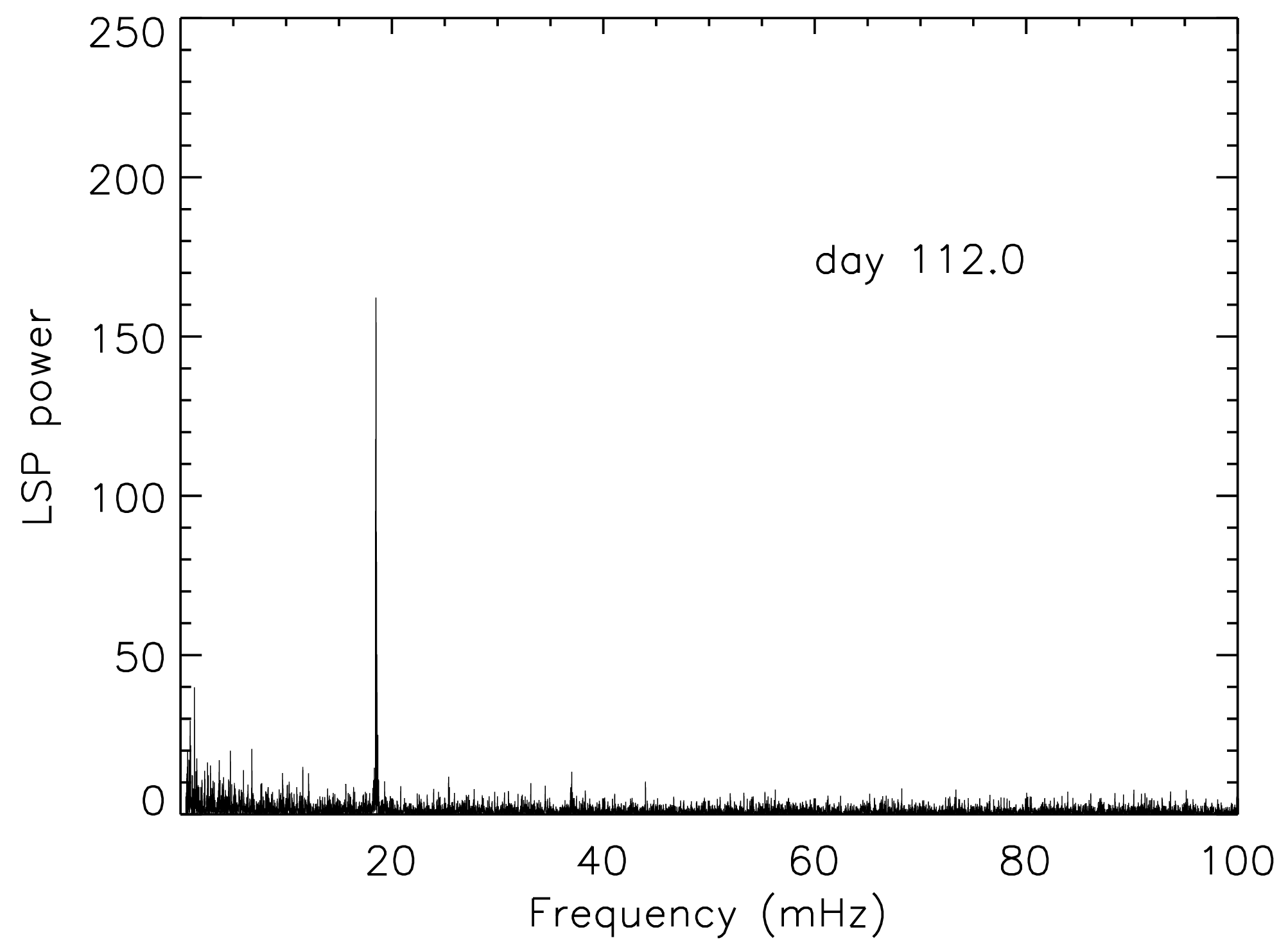}
\hspace{-0.62em}
\includegraphics[width=88mm]{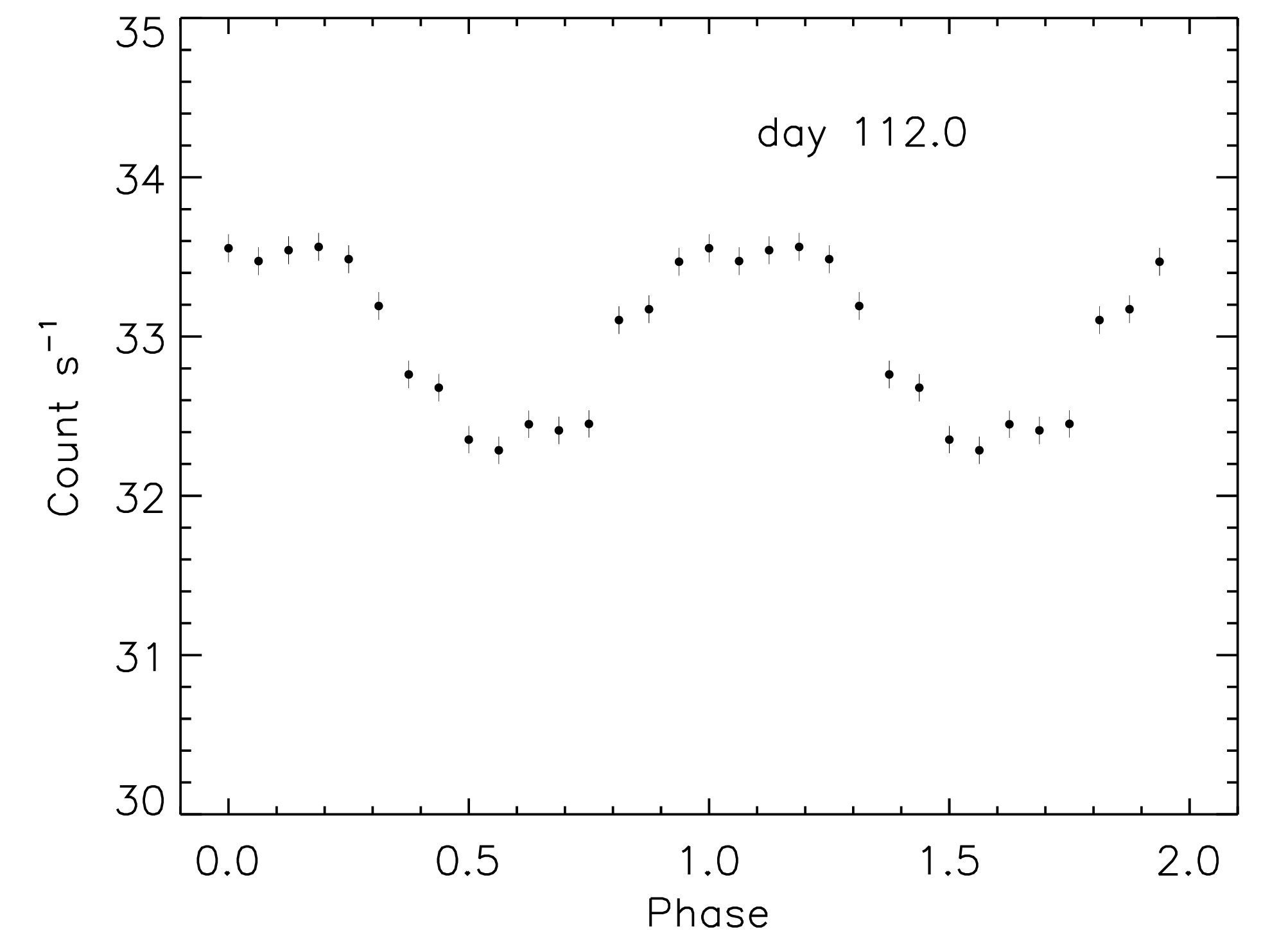}

\caption{Left: Lomb-Scargle periodogram of the light curve from the second {\it Chandra} observation 15743 of V339 Del. Right: The light curve folded with the detected period of 54.08 seconds (corresponding to $\simeq$ 18.49 mHz). The epoch of the first data point is used as the zero point.}
\label{fig: lsp and pfold lc ch}
\end{center}
\end{figure*}
\begin{figure*}
\centering
\includegraphics[width=17.0cm]{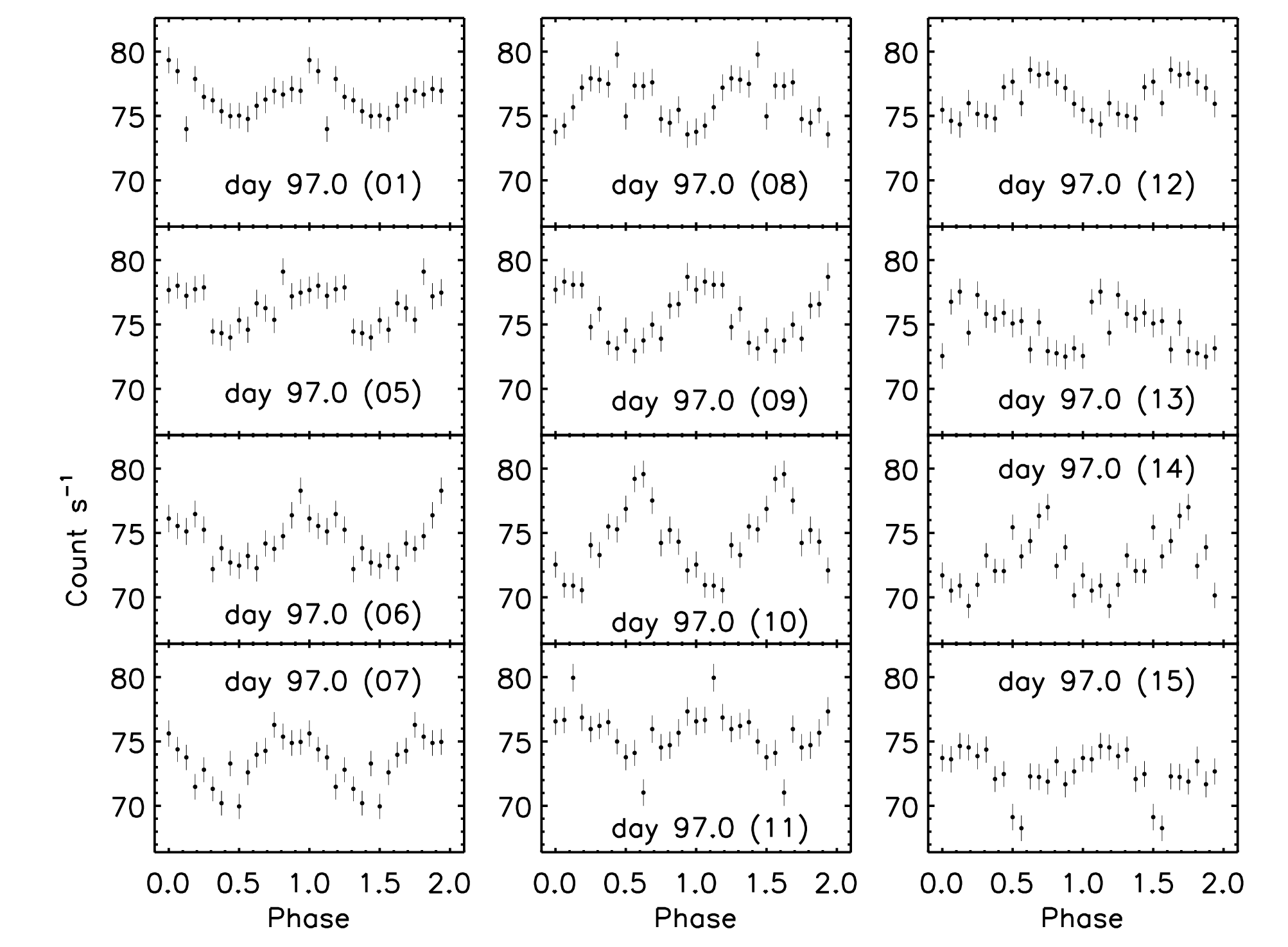}
\caption{Time evolution of phase-folded light curves (MOS 2) from the {\it XMM-Newton} observation 0728200201 of V339 Del on day 97.0. The 12 panels, arranged from top to bottom and left to right, represent light curves for the 12 subintervals (2040 seconds per subinterval), folded using the 54.06-second period. The epoch of the first data point is set as the zero point. In each panel, the x-axis displays the phase over two cycles for enhanced visibility.}\label{fig:folded light curves 0728}
\end{figure*}
\begin{figure*}
\centering
\includegraphics[width=17.0cm]{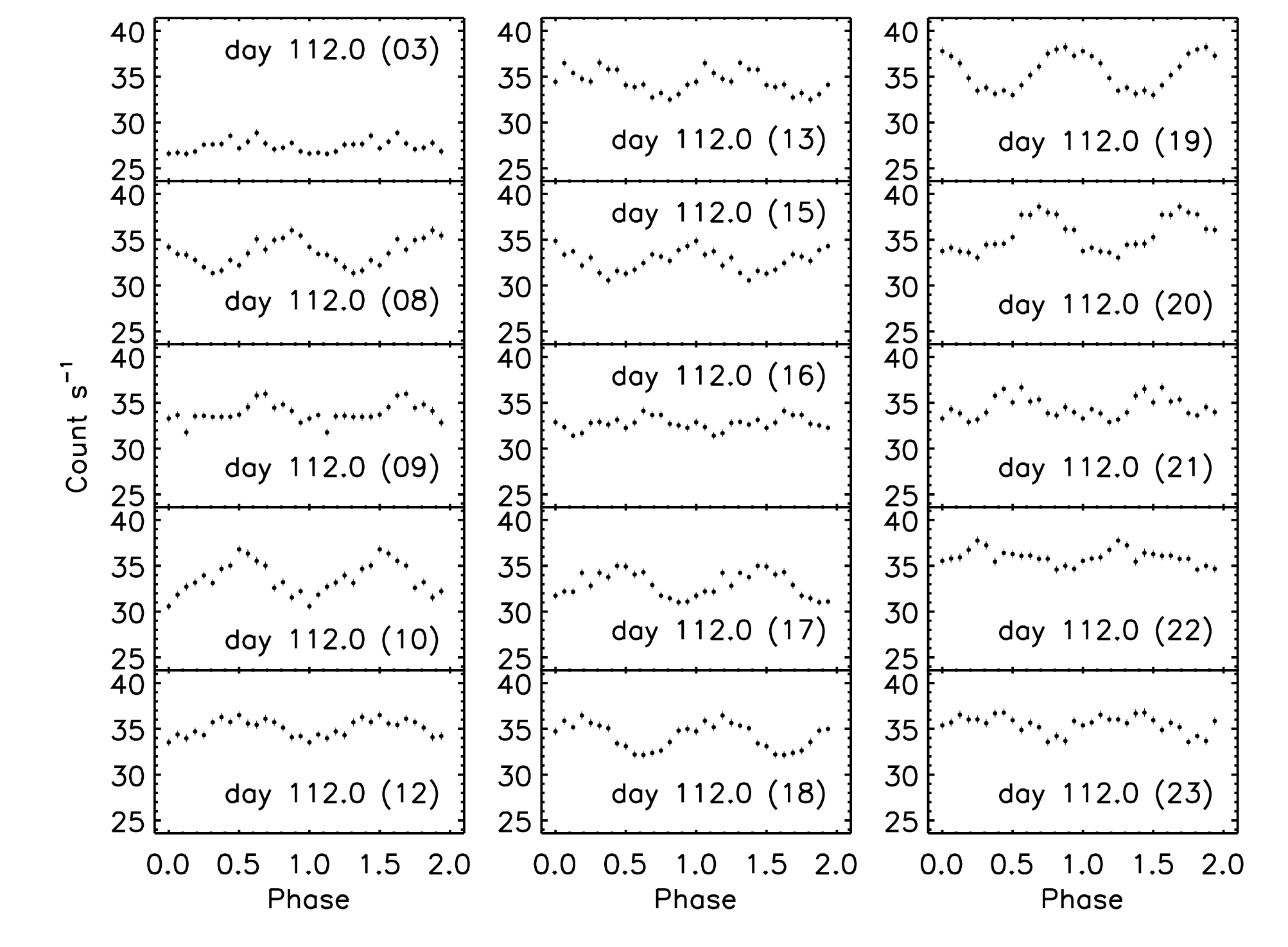}
\caption{Time evolution of phase-folded light curves (HRC-S) from the {\it Chandra} observation 15743 of V339 Del on day 112.0. The 15 panels, arranged from top to bottom and left to right, represent light curves for the 15 subintervals (2040 seconds per subinterval), folded using the 54.08-second period. The epoch of the first data point is set as the zero point. In each panel, the x-axis displays the phase over two cycles for better visibility.}\label{fig:folded light curves 15743}
\end{figure*}
\begin{figure*}
\centering
\includegraphics[width=17.0cm]{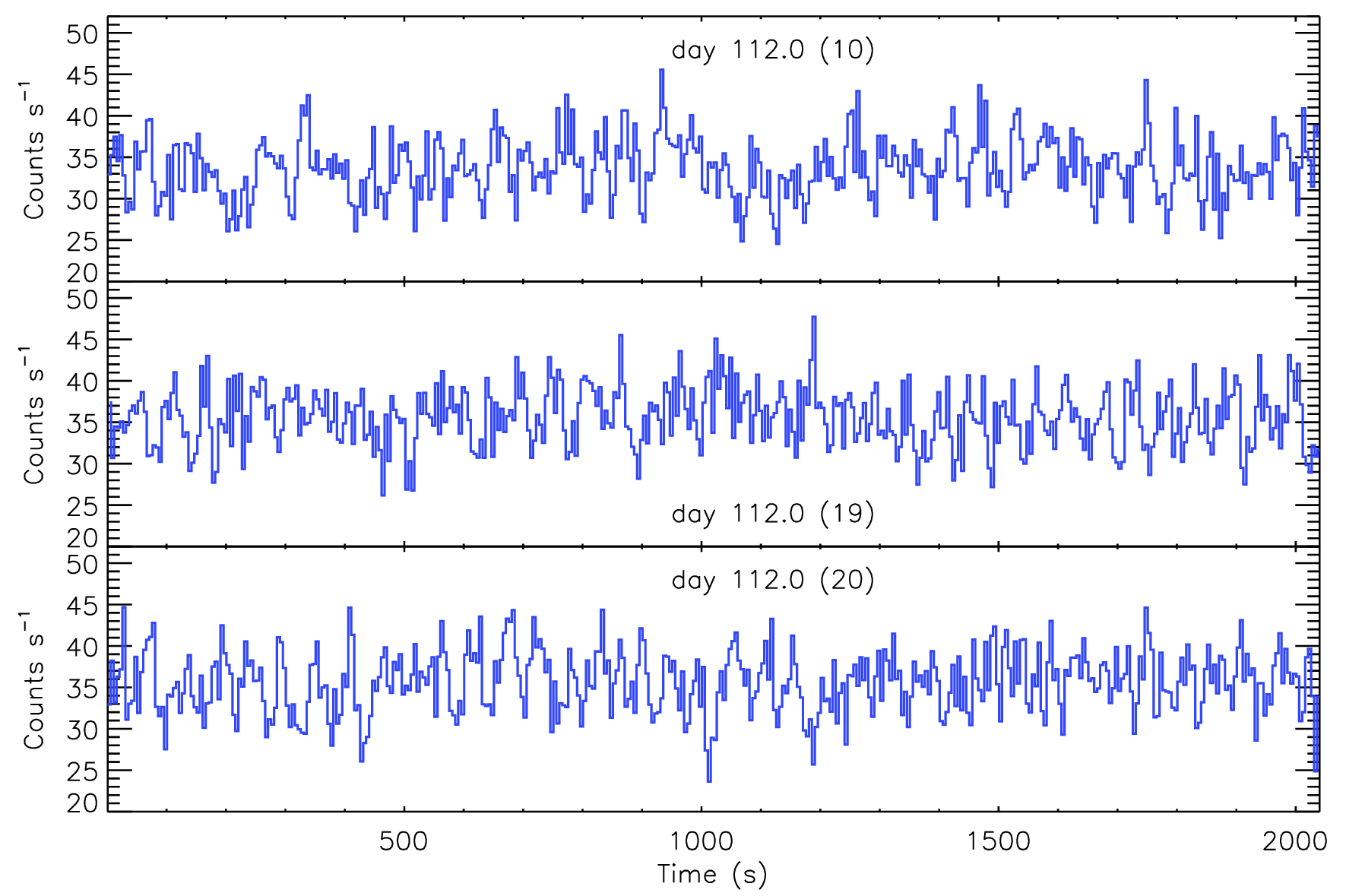}
\caption{Three examples demonstrate how the $\sim$ 54-second period modulation is visually evident. From top to bottom, the three panels show 5-second-per-bin light curves for the tenth, nineteenth, and twentieth subintervals of the second {\it Chandra} observation on day 112.0.}\label{fig:three examples lc with period}
\end{figure*}
\begin{figure*}
\centering
\includegraphics[width=17.0cm]{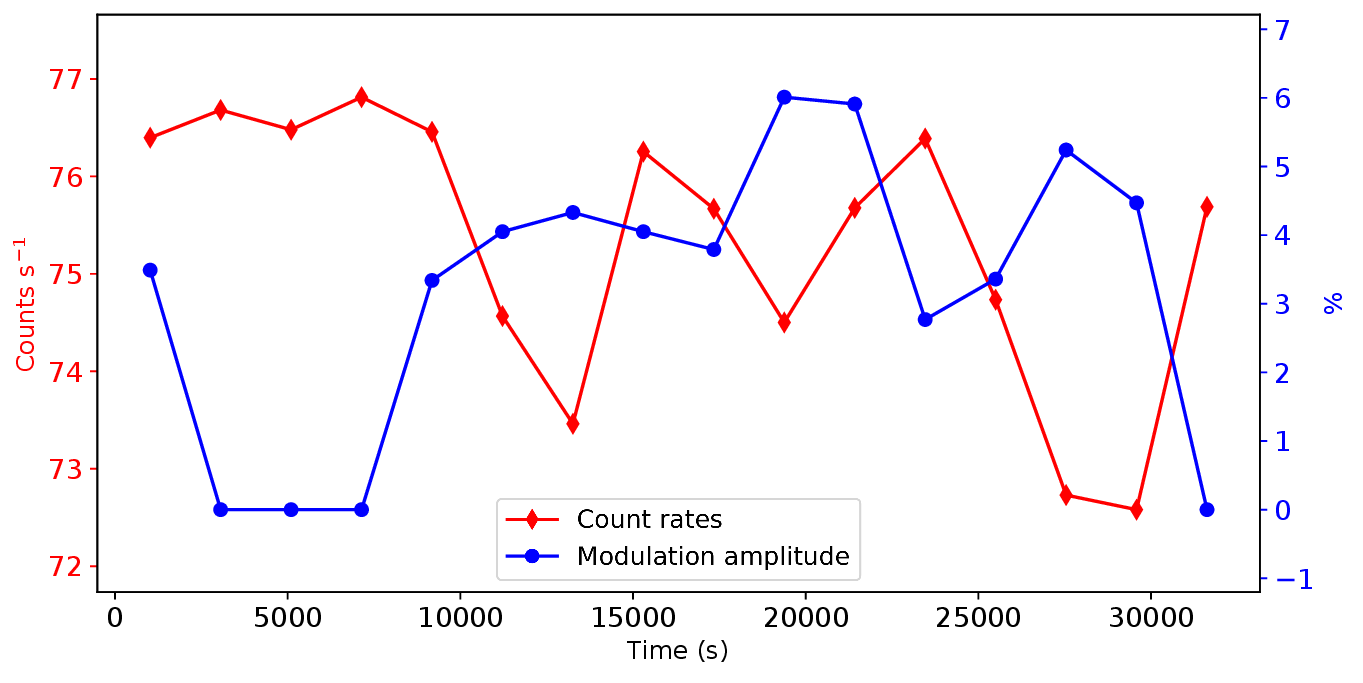}
\caption{The MOS 2 light curve (red diamonds) of V339 Del, measured with {\it XMM-Newton} on day 97.0, is binned every 2040 seconds. It is displayed with time on the x-axis and count rate on the y-axis (left panel). The modulation amplitudes of the $\sim$ 54-second period (blue circles) for the 16 subintervals (each 2040 seconds) derived from the MOS 2 light curve (see Table~\ref{tab:Period}) are shown with time on the x-axis and modulation amplitude on the y-axis (right panel).}\label{fig:apm acr}
\end{figure*}
\begin{figure*}
\centering
\includegraphics[width=17.8cm]{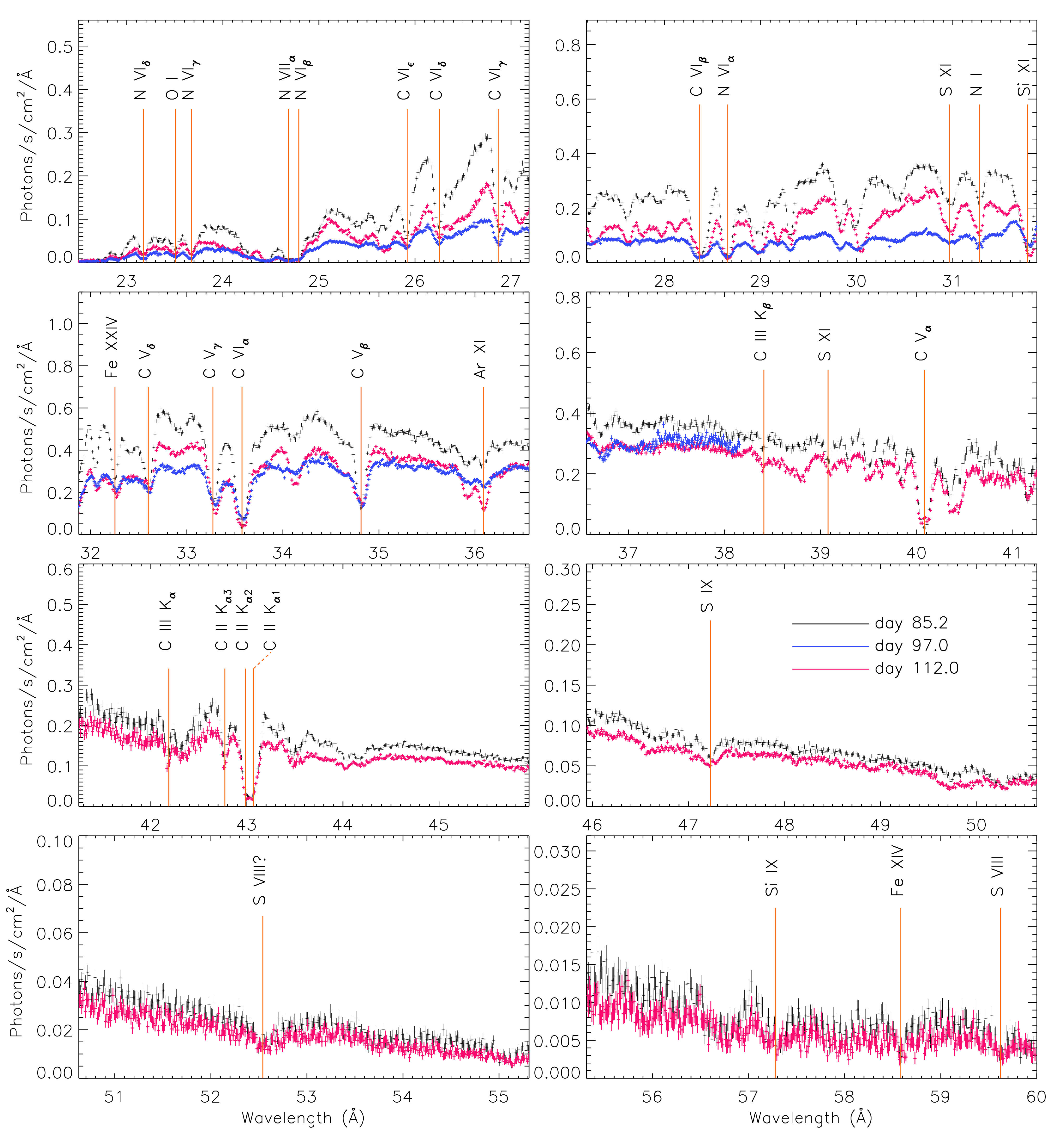}
\caption{The X-ray grating spectra of V339 Del, measured with {\it Chandra} and {\it XMM-Newton} on days 85.2 (black), 97.0 (blue), and 112.0 (pink) after optical maximum, are shown. Absorption lines identified in the spectra are marked in orange with their proposed identifications, and the corresponding line labels are positioned at the observed wavelengths in the day 85.2 spectrum. The spectra have been binned for clarity, with binning achieved at least 20, 10, and 20 counts per bin for days 85.2, 97.0, and 112.0, respectively.}\label{fig:3spectra}
\end{figure*}
\begin{figure*}
\centering
\includegraphics[width=17.0cm]{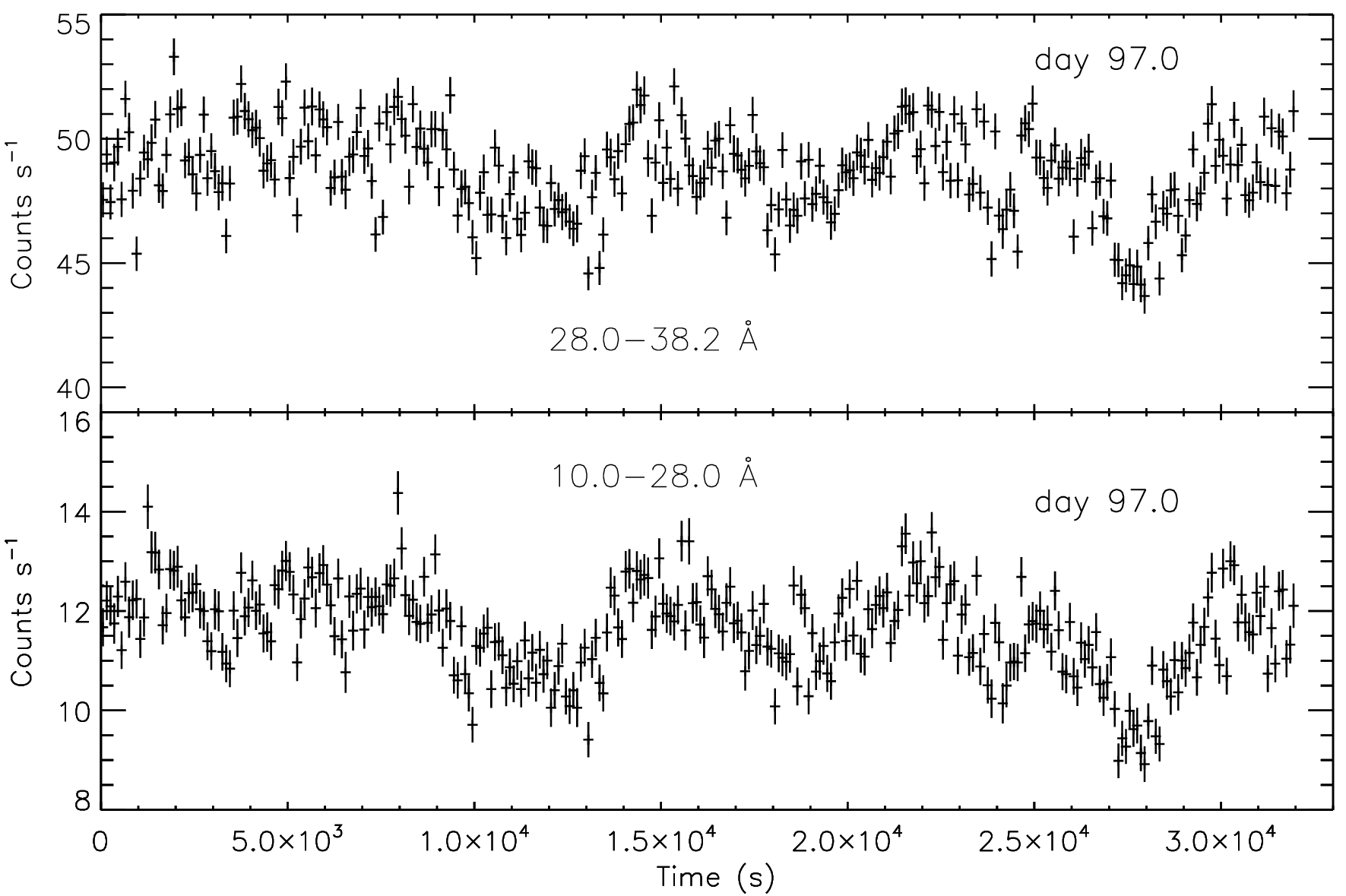}
\caption{The background-subtracted zero-order light curves of V339 Del, measured with the {\it XMM-Newton} RGS on day 97.0, are binned every 100 seconds. The top panel displays the soft light curve in the 28.0 $-$ 38.2 \AA\ range, while the bottom panel shows the hard light curve in the 10.0 $-$ 28.0 \AA\ range. Note that different y-axis scales are used for the two panels.}\label{fig:soft and hard lc}
\end{figure*}
\begin{figure}
\begin{center} 
\includegraphics[width=0.98\columnwidth]{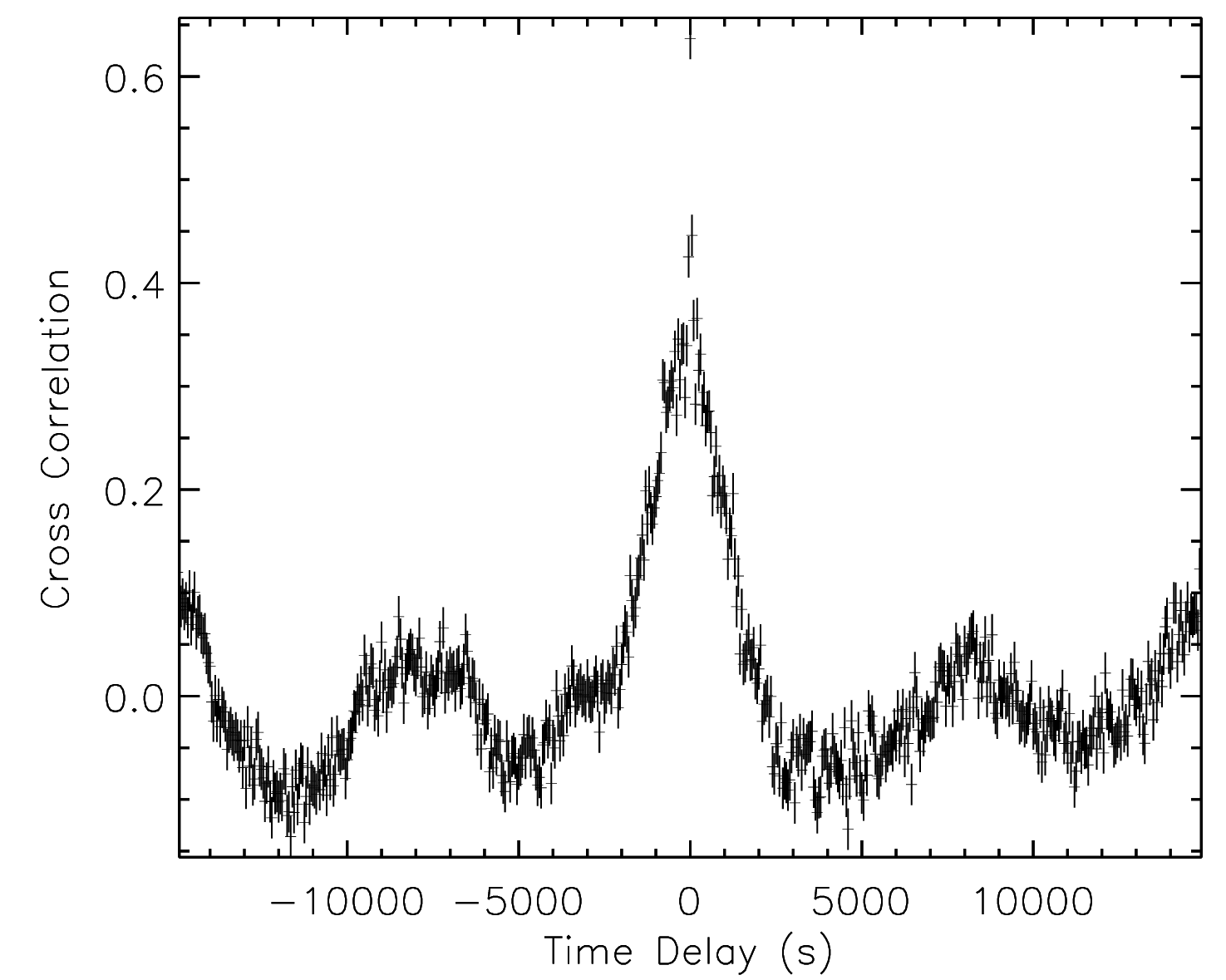}
\caption{The cross-correlation functions, computed with time bin of 48 seconds, of the soft and hard light curves shown in Fig.~\ref{fig:soft and hard lc} which are binned every 10 seconds; see Sect. \ref{sec:Appearance} for details.}
\label{fig:ccf}
\end{center}
\end{figure}
\begin{figure*}
\begin{center}
\includegraphics[width=8.5cm]{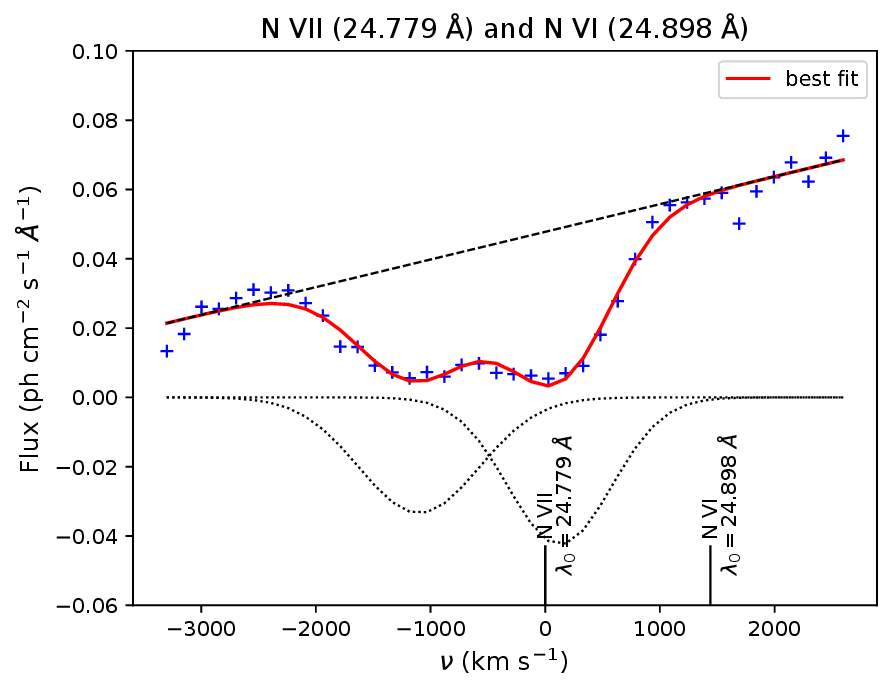}
\hspace{0.08em}
\includegraphics[width=8.5cm]{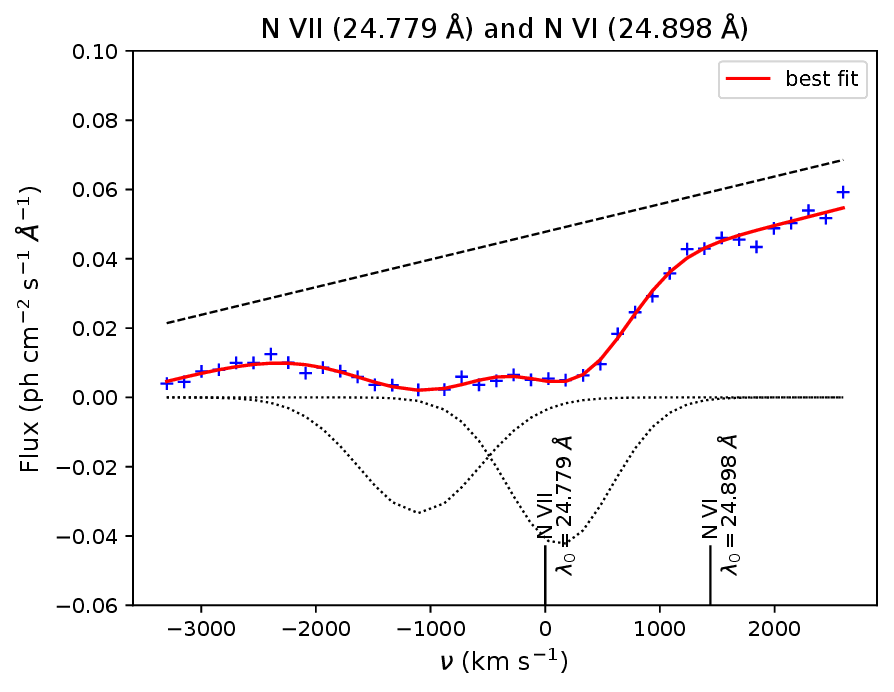}
\caption{Fit to the blended lines of N VII ($\lambda_0=24.779$\,\AA) and N VI ($\lambda_0=24.898$\,\AA) in the spectra of V339 Del on days 85.2 (left) and 112.0 (right). The blue crosses represent the observed flux with error bars, the solid red lines indicate the best fit using the method described by \citet{2010AN....331..179N} and \citet{2011ApJ...733...70N}, and the dotted black lines and dashed black line represent the Gaussian-like component and the continuum component in the spectra, respectively. The parameters are provided in Table~\ref{table:absorption V339}.}\label{fig: line fits 1}
\end{center}
\end{figure*}
\begin{figure*}
\begin{center}
\includegraphics[width=5.6cm]{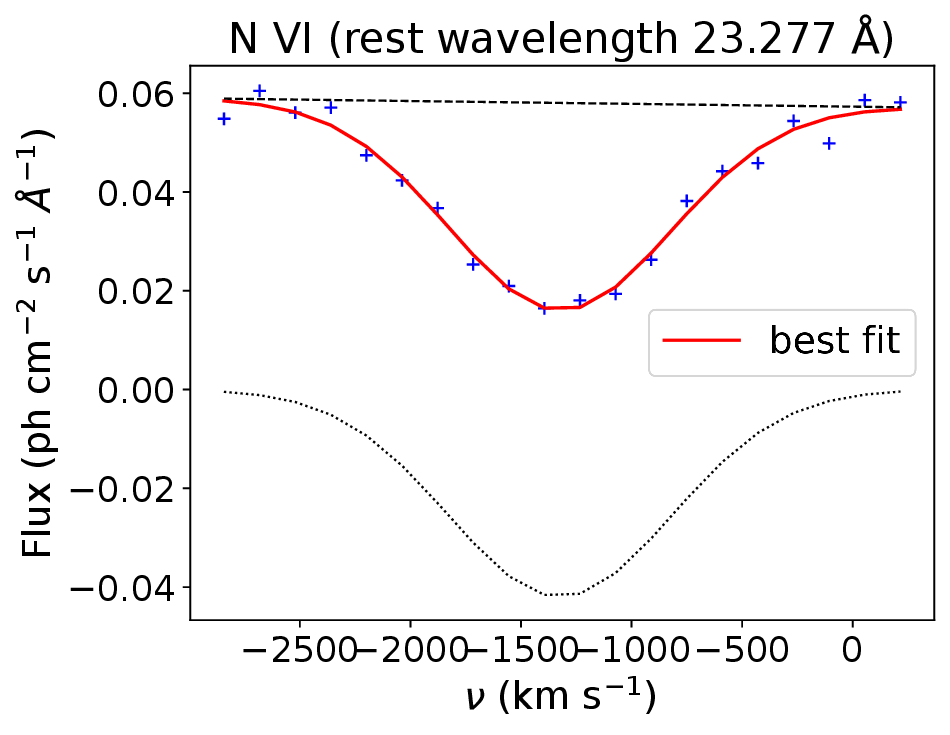}
\hspace{-0.01em}
\includegraphics[width=5.6cm]{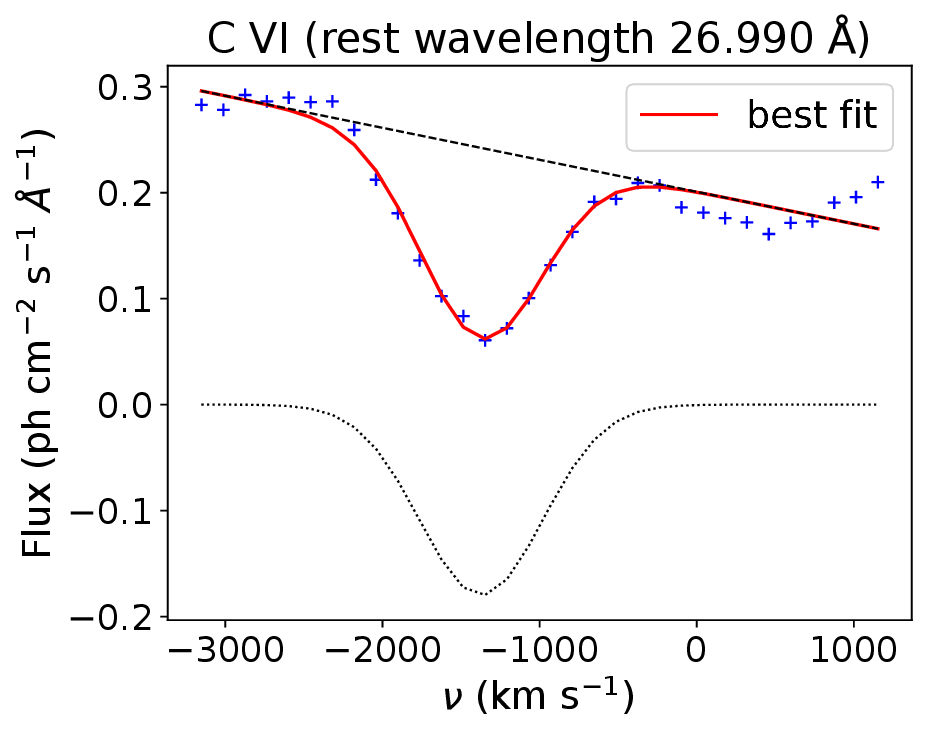}
\hspace{-0.01em}
\includegraphics[width=5.6cm]{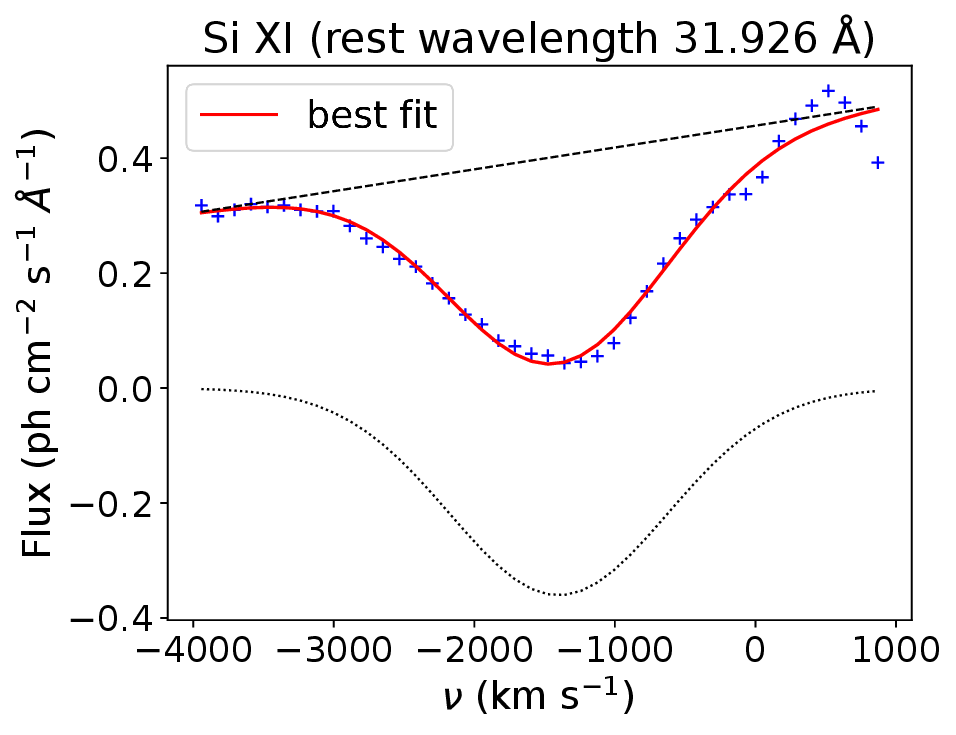}
\caption{Fit to the absorption lines of N VI ($\lambda_0=23.277$\,\AA), C VI ($\lambda_0=26.990$\,\AA), and Si XI ($\lambda_0=31.926$\,\AA) on day 85.2. The blue cross symbols represent the observed spectra with error bars, the solid red lines indicate the best fit using the method described by \citet{2010AN....331..179N} and \citet{2011ApJ...733...70N}, and the dotted black lines and dashed black line represent the Gaussian-like component and continuum component in the spectra, respectively. The parameters are provided in Table~\ref{table:absorption V339}.}\label{fig: line fits 2}
\end{center}
\end{figure*}
\begin{figure*}
\centering
\includegraphics[width=17.0cm]{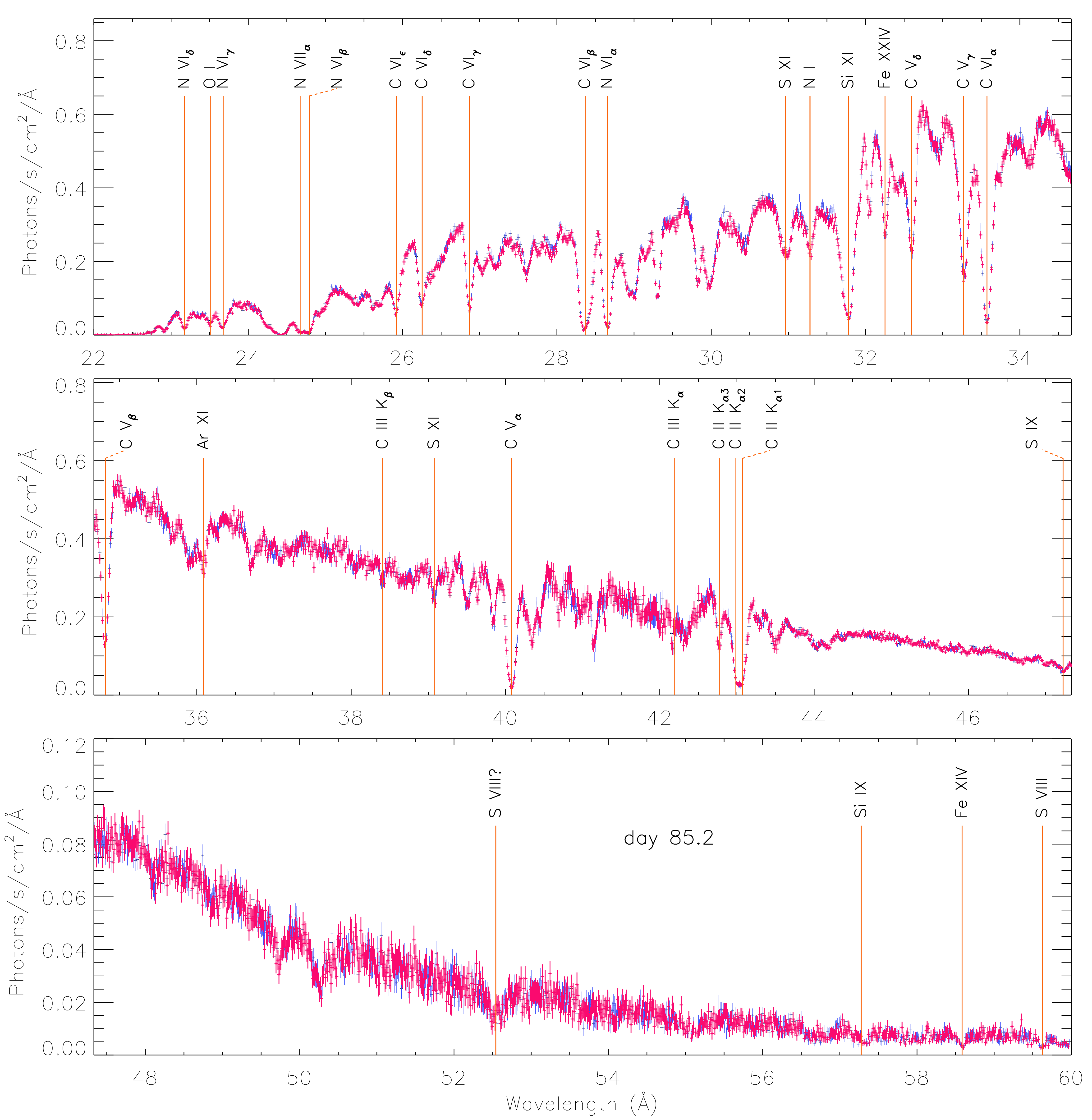}
\caption{Comparison of the `high count rate' and `low count rate' LETG spectra of V339 Del on day 85.2 after the optical-maximum. The green crosses represents the spectrum extracted when the zero-order light curve count rate exceeded the average of 47.18 counts s$^{-1}$ measured during the exposure. The pink crosses represents the spectrum extracted when the count rate was below this average. The spectra have been binned to achieve at least 20 counts per bin for better clarity. Absorption lines identified in the spectra are marked in orange with their proposed identifications, and the corresponding line labels are positioned at the observed wavelengths in the day 85.2 spectrum.}\label{fig:2spectra01}
\end{figure*}
\begin{figure*}
\centering
\includegraphics[width=17.0cm]{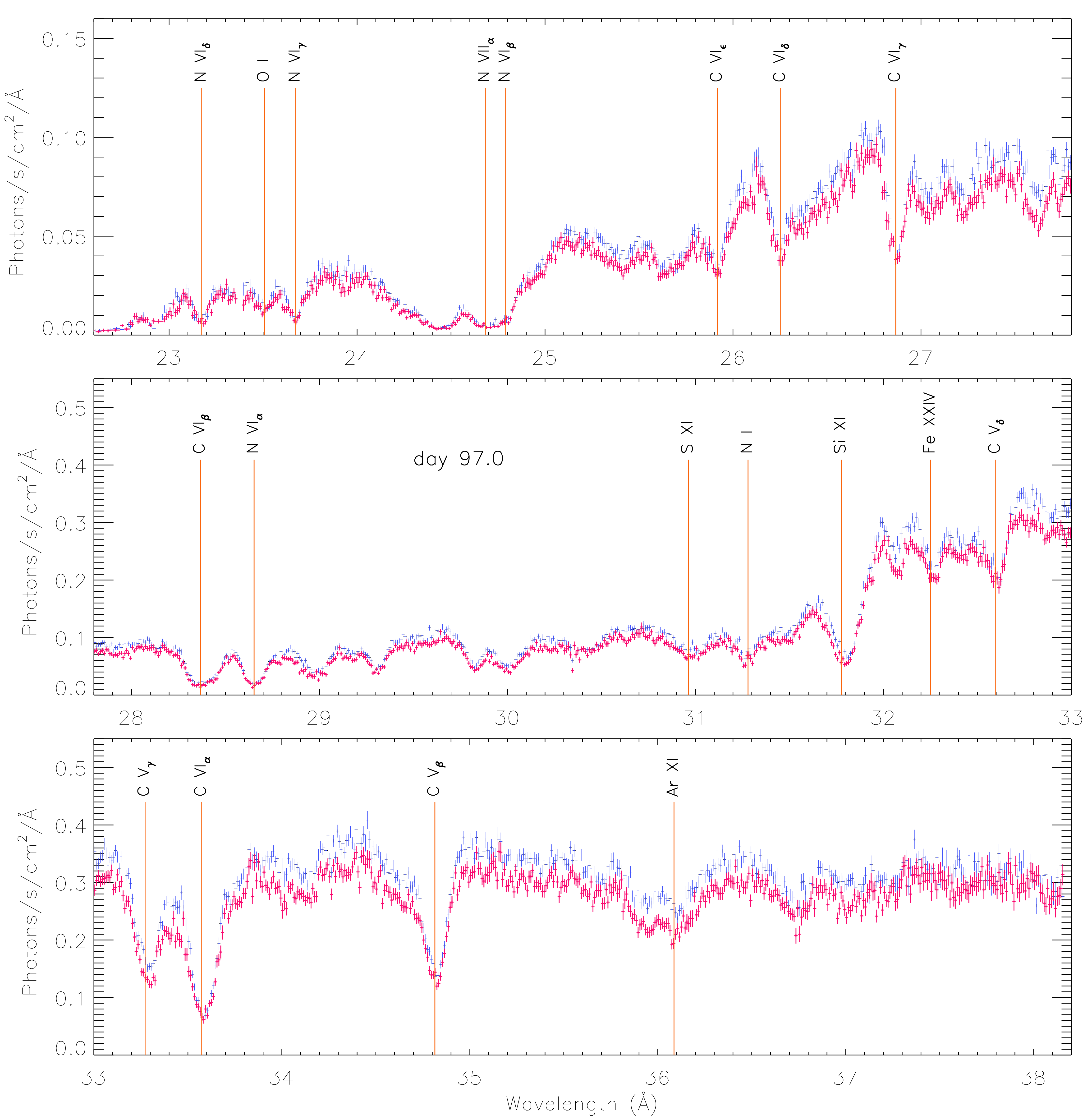}
\caption{Comparison of the `high count rate' and `low count rate' RGS spectra of V339 Del on day 97.0 after the optical-maximum. The green crosses represents spectrum extracted when the zero-order light curve count rate was above the average of 62.88 counts s$^{-1}$ measured during the exposure. The pink crosses represents spectrum extracted when the count rate was below this average. The spectra have been binned to ensure at least 10 counts per bin for better clarity. Absorption lines identified in the spectra are marked in orange with their proposed identifications, and the corresponding line labels are positioned at the observed wavelengths in the day 85.2 spectrum.}\label{fig:2spectra02}
\end{figure*}
\begin{figure*}
\centering
\includegraphics[width=17.0cm]{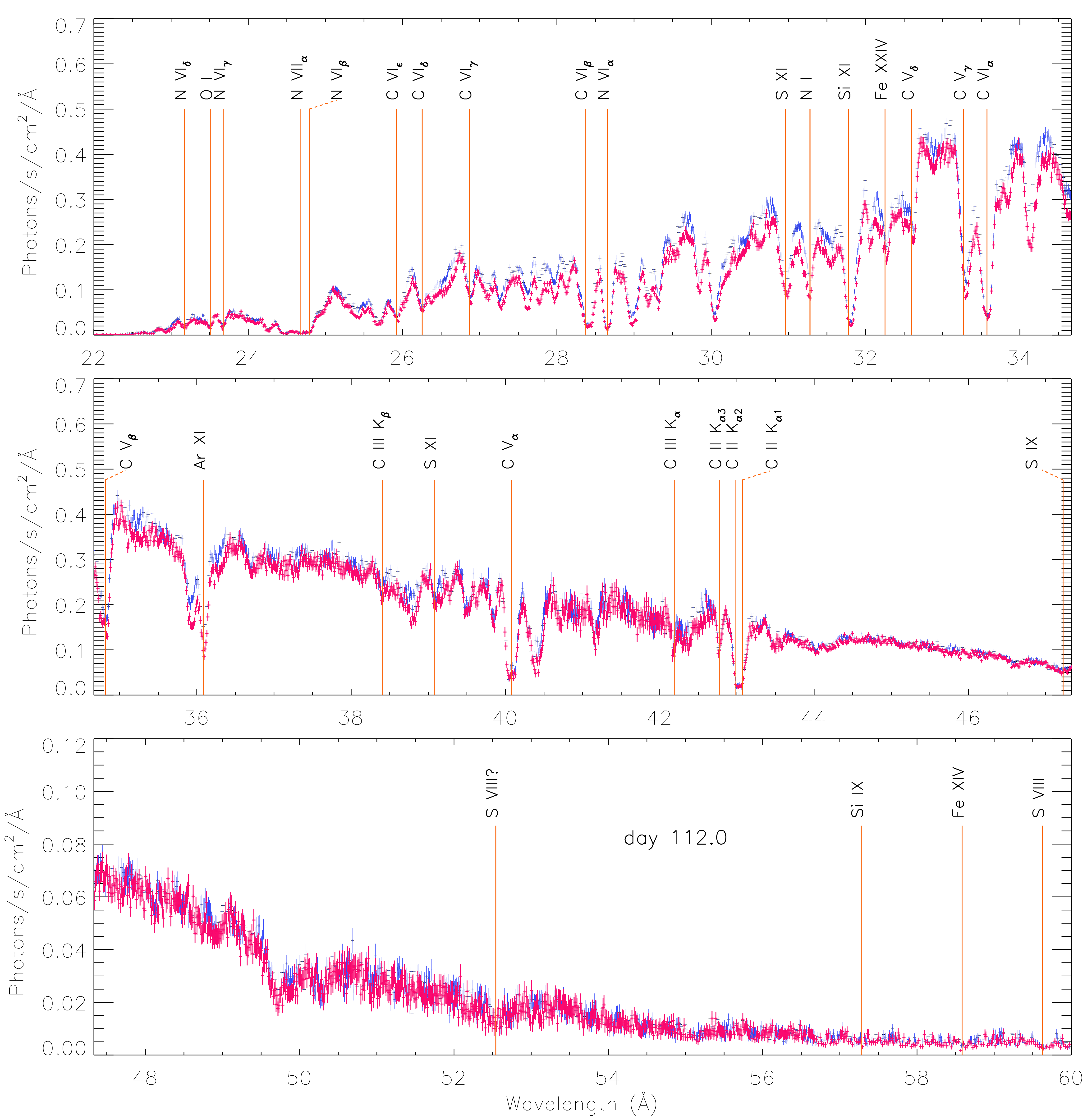}
\caption{Comparison of the `high count rate' and `low count rate' LETG spectra of V339 Del on day 112.0 after the optical-maximum. The green crosses represents spectrum extracted when the zero-order light curve count rate was above the average of 33.00 counts s$^{-1}$ measured during the exposure. The pink crosses represents spectrum extracted when the count rate was below this average. The spectra have been binned to achieve at least 20 counts per bin for better clarity. Absorption lines identified in the spectra are marked in orange with their proposed identifications, and the corresponding line labels are positioned at the observed wavelengths in the day 85.2 spectrum.}\label{fig:2spectra03}
\end{figure*}

\section{Timing analysis}
\label{sec:timing}
Fig.~\ref{fig:lc} shows the background-subtracted zero-order light curves measured with the HRC-S camera and the background-subtracted light curve measured with the RGS, binned every 100 seconds for clarity. In Fig.~\ref{fig:sum}, we present the AAVSO optical light curve in both the visual and V bands, along with the {\it Swift} XRT light curves in different energy bands and the corresponding X-ray hardness ratio. The {\it Swift} XRT exposures ranged from a few tens of seconds to $\sim$ 15 ks per ObsID.

The X-ray light curves obtained on days 85.2, 97.0, and 112.0 show mild variability. The count rates fluctuate during all three {\it Chandra} and {\it XMM-Newton} observations over relatively short time scales, with the average count rate also varying between different exposures.

\subsection{Emergence and disappearance of a short periodic modulation}
\label{sec:Appearance}
A period of 53.2 $\pm$ 1.2 seconds was initially detected in the {\it Swift} XRT light curve by \citet{2013ATel.5573....1B} and later confirmed in the {\it XMM-Newton} light curve with a period of 54.06 seconds \citep{2013ATel.5626....1N}. \citep{2015A&A...578A..39N} analyzed the same {\it Chandra} and {\it XMM-Newton} observation data of V339 Del and found that the period appeared to be variable in both observation epochs, days 97.0 and 112.0 (see their Fig. 7).

Of the two {\it Chandra} observations, the period was only detected in the second observation, on day 112.0. We confirm this detection, identifying a period of $54.06 \pm 0.02$ seconds with 99.0\% significance in the day 97.0 RGS and MOS light curves, and a period of $54.08 \pm 0.01$ seconds with 99.0\% significance in the day 112.0 light curve. Since MOS has higher time resolution than RGS, we focused on the MOS light curve for the timing analysis.

We employed the Lomb-Scargle method \citep{1982ApJ...263..835S} for our analysis. Before applying the Lomb-Scargle method, we detrended the light curves by subtracting the mean and normalizing by the standard deviation. We calculated the Lomb-Scargle periodogram (LSP) using the Starlink PERIOD package. The SCARGLE task was utilized to generate the LSP for each light curve by scanning the frequency space. The resulting LSP frequency plots are shown in the left panels of Fig.~\ref{fig: lsp and pfold lc xmm} and Fig.~\ref{fig: lsp and pfold lc ch}. We identified the highest peak in the periodogram using the PERIOD task PEAKS within the specified frequency range. To assess the statistical significance of the detected period and estimate the associated error, we performed a Fisher randomization test \citep{Nemec1985}, including red noise in the significance analysis, across frequencies from 10 to 100 mHz. The period was detected in the light curves from the {\it XMM-Newton} observation and the second {\it Chandra} observation. To address the possibility of missing a short-lived period, we segmented the light curves from the first {\it Chandra} observation into 2000-second ($\approx 37$ cycles of a 54-second period) and 1000-second ($\approx 18.5$ cycles of a 54-second period) intervals. No periodicity was found in any segment of the first {\it Chandra} exposure, consistent with the results of \citet{2013ATel.5593....1N}. The results are summarized in Table~\ref{tab:Period}. Fig.~\ref{fig: lsp and pfold lc xmm} and Fig.~\ref{fig: lsp and pfold lc ch} display the LSPs of the light curves from days 97.0 and 112.0, extracted without energy filtering, alongside the corresponding folded light curves. The period of approximately 54 seconds remains consistent when accounting for errors, although the modulation amplitude varies significantly. For the {\it XMM-Newton} and second {\it Chandra} observations, dividing the light curves into 2000-second segments resulted in 16.3 and 24.5 segments, respectively, which are non-integer values. To address this, we used 2040-second segments ($\approx 37.8$ cycles of a 54-second period), yielding 16 and 24 segments, respectively. A period of approximately 54 seconds was detected in 12 and 15 segments of these exposures, respectively, as detailed in Table~\ref{tab:Period}. This analysis was repeated with different bin sizes, yielding consistent results. The light curves, folded with periods of 54.06 seconds and 54.08 seconds for the 12 and 15 segments respectively, are shown in Fig.~\ref{fig:folded light curves 0728} and Fig.~\ref{fig:folded light curves 15743}. Fig.~\ref{fig:three examples lc with period} illustrates the distinct modulation observed.
 
Table~\ref{tab:Period} shows that the periods of approximately 54 seconds retrieved from the 2040-second intervals are consistent with those obtained from the full exposures, except for three segments. These segments are the first and thirteenth of the {\it XMM-Newton} observation and the sixteenth of the second {\it Chandra} observation. In these cases, the corresponding peaks in the periodogram are less prominent, and the significance levels are somewhat lower for two of these segments. This suggests that there may be slight variations in the 54-second period. This variation has been previously reported by \citet{2015A&A...578A..39N}, who observed a more pronounced period drift than what we find in our results. This discrepancy could be attributed to their use of shorter time intervals (1000 s) for the analysis, compared to the 2040 s intervals employed in our study. Similar period drift phenomena have been observed in other sources, such as the close binary supersoft source CAL 83 \citep{2014MNRAS.437.2948O, 2022ApJ...932...45O}, nova LMC 2009a \citep{Orio2021}, Nova Her 2021 \citep{2021ApJ...922L..42D}, and the 2021 outburst of the nova RS Oph \citep{2021ATel14901....1P, 2023ApJ...955...37O}, where the period drift decreased towards the end of the supersoft source phase, leading to a more stable period \citep{2023ApJ...955...37O}.

The phase-folded light curve across the three observations reveals significant variations in the pulse profile (see Fig.\ref{fig: lsp and pfold lc xmm}, Fig.\ref{fig: lsp and pfold lc ch}, Fig.\ref{fig:folded light curves 0728}, and Fig.\ref{fig:folded light curves 15743}), which is also evident in the light curve shown in Fig.~\ref{fig:three examples lc with period}. The pulse profiles for the complete observations on days 97.0 and 112.0 are smoother compared to those of the 2040-second segments, likely due to the longer integration time for the full observations. However, the pulse profiles for the entire observations on days 97.0 and 112.0 exhibit slightly broader or less symmetric structures compared to a perfect sinusoidal curve. Similarly, the individual pulse profiles for the 2040-second segments are not consistently sinusoidal. These deviations from a purely sinusoidal shape may be related to period drift. Similar pulse profile characteristics have been observed in Nova Her 2021 \citep{2021ApJ...922L..42D} and RS Oph \citep{2023ApJ...955...37O}. For example, the phase-folded light curve of Nova Her 2021 shows significant deviations from a pure sine wave (see Fig. 2 in \citealt{2021ApJ...922L..42D}), while the phase-folded light curve of RS Oph during its 2021 outburst demonstrates subtle departures from a pure sine wave (see Fig. 19 in \citealt{2023ApJ...955...37O}).

The modulation amplitude of the approximately 54-second period for the entire observation on day 112.0 is lower than that for day 97.0. However, the average modulation amplitude for the 2040-second segments on day 112.0 is higher than that on day 97.0. This difference is attributed to the lower duty cycle of the 54-second period on day 112.0 compared to day 97.0, with the 54-second period being present during 62.5\% of the observation time on day 112.0, compared to 75\% on day 97.0. For the 2040-second segments where the 54-second period is present, the modulation amplitude varies over time (see Table~\ref{tab:Period}). As shown in Fig.~\ref{fig:apm acr}, the modulation amplitudes on day 97.0 exhibit an anti-correlation with count rates, whereas no such anti-correlation is observed on day 112.0. A similar pattern was noted in the 2021 outburst of the recurrent symbiotic nova RS Ophiuchi, where the period power of the 35-second period showed a possible anti-correlation with count rates during the latter part of the {\it XMM-Newton} observation on day 55.6 after the optical peak \citep{2023A&A...670A.131N}. The reason for this anti-correlation remains unexplained. The presence or absence of the 54-second period in the X-ray light curve, along with its varying modulation amplitude over time scales of a few thousand seconds, likely indicates that emission from the central hot source was affected by temporary obscuration events (see discussion in Sect. \ref{sec:broad}). On day 85.2, the count rate varied by a factor of 1.20 during the exposure. On day 97.0, it varied by a factor of 1.25, and on day 112.0, by a factor of 1.62. For segments 8 to 24 on day 112.0, the variation was by a factor of 1.31. The pulsations with a 54-second time scale appeared in the X-ray light curve when the count rate varied by factors between 1.25 and 1.31. This is similar to the case of KT Eridani, where the 35-second period was clearly visible more than 100 days after optical maximum, with the count rate varying by a factor of 1.40 \citep{2021MNRAS.507.2073P}.

We did not detect any additional periodicities in any of the three observations. The proposed periods of 3.154 hours \citep{2015gacv.workE..56C}, 6.43 hours \citep{2014CoSka..43..330C}, and $3.910584 \pm 0.001440$ hours \citep{2022MNRAS.517.3640S}, which were identified in optical light curves, are too long to be measured within the durations of the {\it Chandra} and {\it XMM-Newton} observations.

Fig.~\ref{fig:3spectra} shows that the flux on day 85.2 is the highest, while the flux on day 97.0 is lower than on day 112.0 in the 22.5 $-$ 35.7 \AA\ range. However, in the 35.7 $-$ 37.0 \AA\ region, the flux on day 97.0 matches that on day 112.0, and in the 37.0 $-$ 38.2 \AA\ range, the flux on day 97.0 even exceeds that on day 112.0. This suggests that the flux on day 97.0 is softer compared to day 112.0. Fig.~\ref{fig:sum} illustrates the variability of the hardness ratio during the SSS phase. To examine the evolution of the soft (28.0 $-$ 38.2 \AA) and hard (10.0 $-$ 28.0 \AA) light curves, we present the data from day 97.0 in Fig.~\ref{fig:soft and hard lc}. Due to very low counts below 10.0 \AA\ and no counts above 38.2 \AA\ in the RGS light curves, we limited the analysis to the 10.0 $-$ 38.2 \AA\ range. The light curves in Fig.~\ref{fig:soft and hard lc} are related to the RGS spectrum shown in Fig.~\ref{fig:3spectra}, which is why the RGS light curves are displayed in Fig.~\ref{fig:soft and hard lc}. Both light curves exhibit similar evolutionary profiles, although the hard energy band shows larger fluctuations: the count rate varies by factors of 1.22 and 1.48 for the soft and hard light curves, respectively. The $\simeq$ 54 s period was detected in both light curves, whereas it was not observed in the ultraviolet (UV) light curves (2120 \AA) \citep{2013ATel.5626....1N}. To determine if the soft and hard light curves originate from the same region, we performed a cross-correlation analysis \citep[see][]{2017Ap&SS.362..118P} to obtain the cross-correlation function (CCF) and time lag. Fig.~\ref{fig:ccf} demonstrates a clear positive correlation between the soft and hard light curves with no detectable time lag, indicating that they may originate from the same region.

\begin{table}
\begin{minipage}{80mm}
\caption{Temporal analysis of the light curves reveals a period of approximately 54 seconds. The reported errors are provided at the 90\% confidence level.}
\label{tab:Period}
\begin{tabular}{lccc}
\hline
\hline

      Observation & Period & Significance & Modulation   \\    
       & (s) & (\%) & amplitude$^a$ (\%)                    \\
\hline

day 97.0$^b$ &$54.06 \pm 0.02$   &99.0  &2.57 \\
day 97.0$^b$ (01)$^c$ &$54.73 \pm 0.37$   &94.0  &3.49 \\
day 97.0$^b$ (05)$^c$ &$54.07 \pm 0.36$   &99.0  &3.34 \\
day 97.0$^b$ (06)$^c$ &$53.70 \pm 0.39$   &99.0  &4.05 \\
day 97.0$^b$ (07)$^c$ &$53.81 \pm 0.36$   &99.0  &4.33 \\
day 97.0$^b$ (08)$^c$ &$54.36 \pm 0.36$   &99.0  &4.05 \\
day 97.0$^b$ (09)$^c$ &$54.00 \pm 0.36$   &99.0  &3.79 \\
day 97.0$^b$ (10)$^c$ &$54.00 \pm 0.36$   &99.0  &6.01 \\
day 97.0$^b$ (11)$^c$ &$54.00 \pm 0.36$   &99.0  &5.91 \\
day 97.0$^b$ (12)$^c$ &$54.36 \pm 0.36$   &99.0  &2.77 \\
day 97.0$^b$ (13)$^c$ &$53.64 \pm 0.36$   &99.0  &3.36 \\
day 97.0$^b$ (14)$^c$ &$54.00 \pm 0.36$   &99.0  &5.24 \\
day 97.0$^b$ (15)$^c$ &$54.36 \pm 0.36$   &99.0  &4.47 \\
day 112.0$^b$ &$54.08 \pm 0.01$   &99.0  &1.94 \\
day 112.0$^b$ (03)$^d$ &$53.96 \pm 0.35$   &97.0  &4.15 \\
day 112.0$^b$ (08)$^d$ &$53.96 \pm 0.35$   &99.0  &6.93 \\
day 112.0$^b$ (09)$^d$ &$54.31 \pm 0.36$   &99.0  &6.22 \\
day 112.0$^b$ (10)$^d$ &$54.31 \pm 0.36$   &99.0  &9.24 \\
day 112.0$^b$ (12)$^d$ &$53.96 \pm 0.35$   &99.0  &4.24 \\
day 112.0$^b$ (13)$^d$ &$53.96 \pm 0.35$   &99.0  &5.82 \\
day 112.0$^b$ (15)$^d$ &$53.96 \pm 0.35$   &99.0  &6.57 \\
day 112.0$^b$ (16)$^d$ &$53.61 \pm 0.35$   &95.0  &4.13 \\
day 112.0$^b$ (17)$^d$ &$53.96 \pm 0.35$   &99.0  &6.03 \\
day 112.0$^b$ (18)$^d$ &$53.96 \pm 0.35$   &99.0  &6.28 \\
day 112.0$^b$ (19)$^d$ &$53.96 \pm 0.35$   &99.0  &7.35 \\
day 112.0$^b$ (20)$^d$ &$53.96 \pm 0.35$   &99.0  &7.78 \\
day 112.0$^b$ (21)$^d$ &$54.31 \pm 0.36$   &99.0  &5.43 \\
day 112.0$^b$ (22)$^d$ &$54.31 \pm 0.36$   &99.0  &4.39 \\
day 112.0$^b$ (23)$^d$ &$54.31 \pm 0.36$   &99.0  &4.57 \\
\hline


\end{tabular}
Notes:\hspace{0.1cm} $^a $: We define the period modulation amplitude as $(max-min)/(max+min)$. $^b $: Time in days after the optical-maximum on 2013-08-16. $^c $: Segments in the exposure of day 97.0. $^d $: Segments in the exposure of day 112.0.
\end{minipage} 
\end{table}

\section{Spectral analysis}
\label{sec:spectral}
\subsection{The high-resolution X-ray grating spectra}
\subsubsection{Broad-band spectral analysis} \label{sec:broad}
In Fig.~\ref{fig:3spectra}, we present the summed +1 and -1 order LETG spectra from the two {\it Chandra} observations alongside the first-order RGS spectrum from the {\it XMM-Newton} observation. The high-resolution X-ray grating spectra of V339 Del exhibit a complex system of absorption lines superimposed on a blackbody-like continuum, with no detectable emission lines. We indicate the prominent detected absorption lines. Fig.~\ref{fig: line fits 1} and Fig.~\ref{fig: line fits 2} detail the profiles and fits of specific lines. Due to very low counts below 22 \AA\ and above 60 \AA\ for the spectra of V339 Del, we limited our spectral analysis to the range of 22 $-$ 60 \AA\ for the LETG spectra and the range of 22 $-$ 38.2 \AA\ for the RGS spectra.

The continuum emission in the range of 23 $-$ 50 \AA\ for the spectra on days 85.2, 97.0, and 112.0 varies significantly. However, the emission levels at the centres of the N VII Ly$\alpha$ ($\lambda_0=24.779$\,\AA), N VI He$\beta$ ($\lambda_0=24.898$\,\AA), C VI Ly$\beta$ ($\lambda_0=28.465$\,\AA), N VI He$\alpha$ ($\lambda_0=28.787$\,\AA), C V He$\beta$ ($\lambda_0=34.973$\,\AA), and the C II and C III K-edge features remain unchanged. Additionally, the emission levels at the centres of the other absorption lines are generally higher when the continuum emission is stronger (see Fig.~\ref{fig:3spectra}).

Fig.~\ref{fig:2spectra01}, Fig.~\ref{fig:2spectra02}, and Fig.~\ref{fig:2spectra03} compare the `high count rate' and `low count rate' high-resolution spectra of V339 Del on days 85.2, 97.0, and 112.0 after the optical maximum, respectively. Subtle differences are observed between the two spectra on days 97.0 and 112.0, while the spectra on day 85.2 are nearly identical. We remind the reader that for the spectra on day 85.2, since they are high-count and low-count spectra, they must be different, the variation seems to be 'grey', and if there are changes at all wavelengths, the changes in each bin can be very small making the spectra appear almost identical. On days 85.2, 97.0, and 112.0, the high-count spectra consistently exhibit higher emission across all wavelengths compared to the low-count spectra. It is possible that the emission from the central hot source may be influenced by temporary obscuration events. We refer the reader to \citet{2023A&A...670A.131N} for intensive studies of spectral obscuration signatures.

\subsubsection{Narrow-band spectral analysis} \label{sec:narrow}
For the spectra on day 97.0, the emission levels in the troughs of the absorption lines of C VI Ly$\delta$ ($\lambda_0=26.357$\,\AA), S XI ($\lambda_0=31.050$\,\AA), Si XI ($\lambda_0=31.926$\,\AA), C V He$\gamma$ ($\lambda_0=33.426$\,\AA), and Ar XI ($\lambda_0=36.244$\,\AA) are lower in the low-count spectrum compared to the high-count spectrum. The emission levels in the troughs of the remaining absorption lines are also lower in the low-count spectrum, but the differences are minimal. The high-count phase provides additional continuum emission, and the emission levels in the troughs of the remaining absorption lines remain almost unchanged.

For the spectra on day 112.0, the emission levels in the troughs of the absorption lines of S XI ($\lambda_0=31.050$\,\AA), N I ($\lambda_0=31.3$\,\AA), C V He$\gamma$ ($\lambda_0=33.426$\,\AA), C V He$\beta$ ($\lambda_0=34.973$\,\AA), and Ar XI ($\lambda_0=36.244$\,\AA) are lower in the low-count spectrum compared to the high-count spectrum. Similar to the day 97.0 spectra, the emission levels in the troughs of the remaining absorption lines are lower in the low-count spectrum, but the differences are minimal. The high-count phase provides additional continuum emission, and the emission levels in the troughs of the remaining absorption lines remain almost unchanged.

The Si XI ($\lambda_0=31.926$\,\AA) absorption line in the high-count spectrum on day 97.0 is less blue-shifted compared to that in the low-count spectrum. However, for the remaining absorption lines across the spectra on days 85.2, 97.0, and 112.0, no significant difference in the blue-shift of the absorption troughs is detectable.

The "W" shape of the profile of the C V He$\alpha$ ($\lambda_0=40.267$\,\AA) absorption line was only observed in the spectrum on day 112.0 \citep{2023ApJ...943...31M}. The reversal of this line is due to the fact that the source function for the line increases outward after passing through a minimum, which occurs at optical depths greater than unity. If the shell is radiation-dominated (photoionized), the rise in the source function is due to resonance scattering of line photons, which dominates over pure absorption and thermal emission \citep{2023ApJ...943...31M}. This "W" shape profile of C V He$\alpha$ ($\lambda_0=40.267$\,\AA) is present in both the high-count and low-count spectra on day 112.0. Additionally, the absorption lines N I ($\lambda_0=31.3$\,\AA) in the spectra on day 97.0 (see Fig.\ref{fig:3spectra} and the low- and high-count spectra in Fig.\ref{fig:2spectra02}), and S IX ($\lambda_0=47.433$\,\AA) in the spectra on day 112.0 (see Fig.\ref{fig:3spectra} and the low- and high-count spectra in Fig.\ref{fig:2spectra03}), also exhibit this "W" shape profile. We propose that the explanation provided by \citet{2023ApJ...943...31M} may also apply to the "W" shape profiles of the N I ($\lambda_0=31.3$\,\AA) and S IX ($\lambda_0=47.433$\,\AA) absorption lines.

\subsection{Identification and blue shift of individual lines}
\label{sec:lines}
The high-resolution X-ray spectra (22 $-$ 60 \AA\ or 22 $-$ 38.2 \AA\ range) obtained from {\it Chandra} and {\it XMM-Newton} on days 85.2, 97.0, and 112.0 reveal a complex system of absorption lines superimposed on a supersoft source continuum, with no clear emission lines observed. Absorption lines due to transitions of nitrogen, carbon, sulfur, silicon, iron, and argon were identified in all three spectra. We employed the method described by \citet{2010AN....331..179N} and \citet{2011ApJ...733...70N} to determine the observed wavelengths ($\lambda_m$), line shifts ($\lambda_{\rm shift}$), widths ($\lambda_{\rm width}$), and optical depths ($\tau_{\rm c}$) at the line centre for each spectrum. The results are detailed in Table~\ref{table:absorption V339}. For the absorption line fits, we used an absorbed blackbody model to represent the continuum component. Representative fits are illustrated in Fig.~\ref{fig: line fits 1} and Fig.~\ref{fig: line fits 2}.

The absorption lines of O I ($\lambda_0=23.508$\,\AA), N I ($\lambda_0=31.28$\,\AA), and C II ($\lambda_0=43.10$\,\AA) are clearly observed at their rest wavelengths, and these lines are typically associated with the local interstellar medium (ISM). The lines of O I and N I were previously noted by \citet{2013ATel.5626....1N}. Interstellar lines of O I and N I are also detected at their rest wavelengths in other novae, including V2491 Cyg, V4743 Sgr \citep{2011ApJ...733...70N}, Nova SMC 2016 \citep{Orio2018}, LMC 2009a \citep{Orio2021}, and KT Eridani \citep{2021MNRAS.507.2073P}. Only the O I interstellar line is present at its rest wavelength in the recurrent nova RS Oph \citep{2007ApJ...665.1334N, 2023A&A...670A.131N}. The deep absorption edges at 38.4 \AA, 42.2 \AA, 42.8 \AA, and 43.0 \AA\ correspond to the C II K-edge and C III K-edge features due to neutral carbon in the ISM. These features, observed at their rest wavelengths, were analyzed in detail by \citet{2018MNRAS.479.2457G} using the Galactic nova V339 Del during its SSS phase as a lamp. The C K-edges, spanning the 38 $-$ 44 \AA\ wavelength range, are also present at their rest wavelengths in novae such as KT Eridani, Sgr 2015b, V4743 Sgr \citep{2018MNRAS.479.2457G}, and Nova SMC 2016 \citep{Orio2018}.

The remaining identified absorption lines are attributed to the WD atmosphere and are all blueshifted. The resulting parameters for the observed wavelength, blueshift velocity, broadening velocity, and optical depth for each line are detailed in Table~\ref{table:absorption V339}. If our line identifications are correct, there is a significant range in blueshift velocities, from $\sim$ 724 to $\sim$ 1474 km s$^{-1}$, indicating an expanding region. These blueshift velocities remain relatively constant across the three observational epochs. Strong absorption lines generally exhibit blueshift velocities around 1200 km s$^{-1}$, a finding also noted by \citet{2013ATel.5593....1N} and \citet{2013ATel.5626....1N}. Most absorption lines display blueshift velocities in the range of $\sim$ 982 to $\sim$ 1474 km s$^{-1}$, with exceptions including C VI Ly$\beta$ ($\lambda_0=28.465$\,\AA) and S XI ($\lambda_0=31.050$\,\AA). In contrast, other novae, such as V2491 Cyg, RS Oph \citep{2010AN....331..179N}, Nova SMC 2016 \citep{Orio2018}, and LMC 2009a \citep{Orio2021}, exhibit narrower ranges of blueshift velocities. Additionally, spectra of V4743 Sgr \citep{2011ApJ...733...70N} and KT Eridani \citep{2021MNRAS.507.2073P} reveal at least three distinct velocity systems.

The line widths also vary significantly, with broadening velocities of the absorption lines ranging from 310 to 1280 km s$^{-1}$-substantially larger than the expected instrumental width and broadening (which should be less than approximately 300 km s$^{-1}$). No systematic trend in line width variations over time is evident. Similar to V2491 Cyg \citep{2011ApJ...733...70N}, the absorption lines in V339 Del may originate from a plasma region (the ejecta) with a significant extension, allowing us to observe a range of expansion velocities and different plasma layers. Most broadening velocities fall within the $\sim$ 483 to $\sim$ 961 km s$^{-1}$ range, which is lower than the $\sim$ 982 to $\sim$ 1474 km s$^{-1}$ range of their blueshift velocities. This contrasts with the recurrent nova V3890 Sgr, where the blueshift velocities are comparable to the broadening velocities \citep{2022A&A...658A.169N}.

Certain absorption lines are notably distinct, showing consistently smaller line shifts. The C VI Ly$\beta$ ($\lambda_0=28.465$,\AA) and S XI ($\lambda_0=31.050$,\AA) lines stand out prominently, exhibiting consistently lower blue-shift velocities compared to other lines. Notably, the blue-shift velocity of the C VI Ly$\beta$ line decreases over time, while the S XI line shows a narrow but irregular variation in blue-shift velocity as time progresses. The N VII ($\lambda_0=24.779$\,\AA) and N VI ($\lambda_0=24.898$\,\AA) lines are blended, a phenomenon also observed in the X-ray grating spectra of KT Eridani \citep{2021MNRAS.507.2073P}. We derived the wavelength, line shifts, widths, and optical depths assuming the lines overlap, as illustrated in Fig.~\ref{fig: line fits 1}. Although the fits are generally acceptable, this blending introduces significant uncertainty.

An absorption feature at $29.3$ \AA \ is present in all three spectra. The Chianti atomic database \citep{1997A&AS..125..149D, 2024ApJ...974...71D} lists a possible candidate, Fe XXV ($\lambda_0=29.4132$\,\AA), with an oscillator strength of 5.580. The line width is consistent with that of the Fe XXIV line at $32.377$ \AA \. However, the Fe XXV line forms at much higher temperatures than the nitrogen lines, and it has not been identified in spectra from other novae. Therefore, this absorption feature might represent an unidentified line, see Table 5 in \citet{2011ApJ...733...70N}. Consequently, we have added a question mark next to the Fe XXV line ($\lambda_0=29.4132$\,\AA) in Table~\ref{table:absorption V339}. Additionally, we have added a question mark next to the S VIII line ($\lambda_0=52.756$\,\AA) with an oscillator strength of 2.086, as it is also possible that this feature corresponds to Si X ($\lambda_0=52.611$\,\AA), which has an oscillator strength of 2.439. However, if this absorption feature observed at $\sim$ $52.5$ \AA \ is Si X line ($\lambda_0=52.611$\,\AA), the blue-shifted velocity of approximately 400 km s$^{-1}$ on day 85.2 is lower than the blue-shifted velocities of other identified absorption lines, making the identification of this feature uncertain.

One absorption feature remains unidentified. This absorption feature at $27.71$ \AA \, listed in Table 5 of \citet{2011ApJ...733...70N}, is measured at $27.6$ \AA \, which is close but not perfectly coincident, accounting for the blue-shift, with unidentified lines observed in RS Oph \citep{2011ApJ...733...70N}.

From the analysis of individual line profiles, we can conclude that there is a predominant bulk velocity component of $\sim$ 1200 km s$^{-1}$ with relatively small variations in velocities (line widths).

\setcounter{table}{2}
\begin{table*}
\begin{minipage}{180mm}
\begin{flushleft}
\caption{Rest wavelength, observed wavelength, velocity, broadening velocity, and optical depth resulting from the fits of the absorption lines with proposed identification. The reported errors are at the 90\% confidence level. The rest wavelengths were obtained from the AtomDB database version 3.0.9 \citep{2001ApJ...556L..91S, 2012ApJ...756..128F, 2020Atoms...8...49F}.}
\label{table:absorption V339}
\begin{center}
\begin{tabular}{lccccc}
\hline
Ion$^a$ & $\lambda_0$ & $\lambda_m$ & $v_{\rm shift}$ & $v_{\rm width}$ & $\tau_{\rm c}$  \\
& (\AA) & (\AA) & (km\,s$^{-1}$) & (km\,s$^{-1}$) &  \\
\hline
\multicolumn{6}{c|}{\bf Day 85.2$^b$}\\
\hline
N VI He$\delta$  & 23.277 & $23.174 \pm 0.002$ & \mbox{ $-1322 \pm 25$} & \mbox{ $714 \pm 41$} & \mbox{ $0.042 \pm 0.002$}  \\
N VI He$\gamma$  & 23.771 & $23.674 \pm 0.002$ & \mbox{ $-1222 \pm 28$} & \mbox{ $752 \pm 49$} & \mbox{ $0.045 \pm 0.002$}  \\
N VII Ly$\alpha$  & 24.779 & $24.688 \pm 0.008$ & \mbox{ $-1100 \pm 99$} & \mbox{ $739 \pm 117$} & \mbox{ $0.033 \pm 0.002$}  \\
N VI He$\beta$ r   & 24.898 & $24.789 \pm 0.006$ & \mbox{ $-1306 \pm 98$} & \mbox{ $638 \pm 73$} & \mbox{ $0.042 \pm 0.003$}  \\
C VI Ly$\epsilon$  & 26.026 & $25.918 \pm 0.003$ & \mbox{ $-1246 \pm 27$} & \mbox{ $656 \pm 47$} & \mbox{ $0.103 \pm 0.003$}  \\
C VI Ly$\delta$   & 26.357 & $26.254 \pm 0.004$ & \mbox{ $-1176 \pm 45$} & \mbox{ $705 \pm 160$} & \mbox{ $0.129 \pm 0.009$}  \\
C VI Ly$\gamma$  & 26.990 & $26.866 \pm 0.002$ & \mbox{ $-1372 \pm 26$} & \mbox{ $553 \pm 37$} & \mbox{ $0.179 \pm 0.009$}  \\
C VI Ly$\beta$  & 28.465 & $28.367 \pm 0.003$ & \mbox{ $-1032 \pm 29$} & \mbox{ $1102 \pm 50$} & \mbox{ $0.265 \pm 0.008$}  \\
N VI He$\alpha$ r  & 28.787 & $28.652 \pm 0.002$ & \mbox{ $-1410 \pm 27$} & \mbox{ $717 \pm 42$} & \mbox{ $0.225 \pm 0.009$}  \\
Fe XXV?  & 29.4132$^c$ & $29.308 \pm 0.002$ & \mbox{ $-1072 \pm 21$} & \mbox{ $446 \pm 31$} & \mbox{ $0.186 \pm 0.006$}  \\
S XI   & 31.050 & $30.964 \pm 0.002$ & \mbox{ $-826 \pm 26$} & \mbox{ $961 \pm 43$} & \mbox{ $0.130 \pm 0.004$}  \\
Si XI  & 31.926 & $31.777 \pm 0.002$ & \mbox{ $-1400 \pm 46$} & \mbox{ $1096 \pm 48$} & \mbox{ $0.360 \pm 0.010$}  \\
Fe XXIV & 32.377 & $32.252 \pm 0.003$ & \mbox{ $-1157 \pm 81$} & \mbox{ $441 \pm 28$} & \mbox{ $0.228 \pm 0.006$}  \\
C V He$\delta$   & 32.754 & $32.598 \pm 0.002$ & \mbox{ $-1427 \pm 35$} & \mbox{ $529 \pm 29$} & \mbox{ $0.274 \pm 0.010$}  \\
C V He$\gamma$   & 33.426 & $33.272 \pm 0.002$ & \mbox{ $-1377 \pm 39$} & \mbox{ $637 \pm 30$} & \mbox{ $0.329 \pm 0.006$}  \\
C VI Ly$\alpha$  & 33.740 & $33.574 \pm 0.001$ & \mbox{ $-1474 \pm 32$} & \mbox{ $698 \pm 29$} & \mbox{ $0.415 \pm 0.007$}  \\
C V He$\beta$ r    & 34.973 & $34.814 \pm 0.002$ & \mbox{ $-1366 \pm 28$} & \mbox{ $567 \pm 30$} & \mbox{ $0.364 \pm 0.007$}  \\
Ar XI  & 36.244 & $36.086 \pm 0.003$ & \mbox{ $-1308 \pm 37$} & \mbox{ $432 \pm 50$} & \mbox{ $0.074 \pm 0.006$}  \\
S XI  & 39.240 & $39.077 \pm 0.003$ & \mbox{ $-1244 \pm 37$} & \mbox{ $379 \pm 50$} & \mbox{ $0.065 \pm 0.006$}  \\
C V He$\alpha$ r   & 40.267 & $40.080 \pm 0.002$ & \mbox{ $-1388 \pm 36$} & \mbox{ $629 \pm 34$} & \mbox{ $0.277 \pm 0.011$}  \\
S IX  & 47.433 & $47.226 \pm 0.004$ & \mbox{ $-1305 \pm 38$} & \mbox{ $428 \pm 43$} & \mbox{ $0.019 \pm 0.002$}  \\
S VIII?   & 52.756  & $52.542 \pm 0.010$ & \mbox{ $-1212 \pm 55$} & \mbox{ $791 \pm 89$} & \mbox{ $0.009 \pm 0.001$} \\
Si IX  & 57.512 & $57.277 \pm 0.024$ & \mbox{ $-1226 \pm 130$} & \mbox{ $890 \pm 311$} & \mbox{ $0.004 \pm 0.002$}  \\
Fe XIV & 58.811 & $58.584 \pm 0.004$ & \mbox{ $-1160 \pm 33$} & \mbox{ $318 \pm 37$} & \mbox{ $0.004 \pm 0.001$}  \\
S VIII & 59.882 & $59.624 \pm 0.008$ & \mbox{ $-1288 \pm 40$} & \mbox{ $309 \pm 52$} & \mbox{ $0.003 \pm 0.001$}  \\
\hline
\multicolumn{6}{c}{\bf Day 97.0$^b$}\\
\hline
N VI He$\delta$  & 23.277 & $23.173 \pm 0.002$ & \mbox{ $-1335 \pm 33$} & \mbox{ $646 \pm 52$} & \mbox{ $0.013 \pm 0.001$}  \\
N VI He$\gamma$   & 23.771 & $23.671 \pm 0.002$ & \mbox{ $-1257 \pm 32$} & \mbox{ $572 \pm 51$} & \mbox{ $0.018 \pm 0.001$}  \\
N VII Ly$\alpha$  & 24.779 & $24.700 \pm 0.005$ & \mbox{ $-1100 \pm 98$} & \mbox{ $638 \pm 101$} & \mbox{ $0.013 \pm 0.001$}  \\
N VI He$\beta$ r   & 24.898 & $24.802 \pm 0.002$ & \mbox{ $-1306 \pm 99$} & \mbox{ $739 \pm 116$} & \mbox{ $0.014 \pm 0.001$}  \\
C VI Ly$\epsilon$  & 26.026 & $25.917 \pm 0.002$ & \mbox{ $-1252 \pm 27$} & \mbox{ $651 \pm 39$} & \mbox{ $0.023 \pm 0.002$}  \\
C VI Ly$\delta$   & 26.357 & $26.248 \pm 0.003$ & \mbox{ $-1244 \pm 30$} & \mbox{ $430 \pm 43$} & \mbox{ $0.036 \pm 0.003$}  \\
C VI Ly$\gamma$  & 26.990  & $26.871 \pm 0.003$ & \mbox{ $-1323 \pm 35$} & \mbox{ $485 \pm 49$} & \mbox{ $0.052 \pm 0.004$}  \\
C VI Ly$\beta$  & 28.465 & $28.368 \pm 0.004$ & \mbox{ $-1016 \pm 41$} & \mbox{ $1071 \pm 85$} & \mbox{ $0.070 \pm 0.004$}  \\
N VI He$\alpha$ r  & 28.787 & $28.661 \pm 0.003$ & \mbox{ $-1310 \pm 34$} & \mbox{ $658 \pm 41$} & \mbox{ $0.066 \pm 0.003$}  \\
Fe XXV?  & 29.4132$^c$ & $29.316 \pm 0.003$ & \mbox{ $-991 \pm 31$} & \mbox{ $527 \pm 53$} & \mbox{ $0.041 \pm 0.003$ } \\
S XI   & 31.050 & $30.968 \pm 0.014$ & \mbox{ $-783 \pm 132$} & \mbox{ $919 \pm 262$} & \mbox{ $0.025 \pm 0.006$}  \\
Si XI  & 31.926 & $31.816 \pm 0.007$ & \mbox{ $-1035 \pm 61$} & \mbox{ $1062 \pm 117$} & \mbox{ $0.167 \pm 0.013$}  \\
Fe XXIV  & 32.377 & $32.258 \pm 0.004$ & \mbox{ $-1102 \pm 69$} & \mbox{ $610 \pm 85$} & \mbox{ $0.064 \pm 0.008$}  \\
C V He$\delta$   & 32.754 & $32.600 \pm 0.003$ & \mbox{ $-1412 \pm 32$} & \mbox{ $363 \pm 47$} & \mbox{ $0.227 \pm 0.023$}  \\
C V He$\gamma$   & 33.426 & $33.298 \pm 0.008$ & \mbox{ $-1144 \pm 68$} & \mbox{ $429 \pm 109$} & \mbox{ $0.181 \pm 0.036$}  \\
C VI Ly$\alpha$  & 33.740 & $33.586 \pm 0.012$ & \mbox{ $-1354 \pm 104$} & \mbox{ $941 \pm 233$} & \mbox{ $0.282 \pm 0.054$}  \\
C V He$\beta$ r   & 34.973 & $34.824 \pm 0.006$ & \mbox{ $-1272 \pm 53$} & \mbox{ $402 \pm 81$} & \mbox{ $0.228 \pm 0.038$}  \\
Ar XI  & 36.244 & $36.121 \pm 0.009$ & \mbox{ $-1016 \pm 73$} & \mbox{ $314 \pm 116$} & \mbox{ $0.075 \pm 0.025$}  \\
\hline
\multicolumn{6}{c|}{\bf Day 112.0$^b$}\\
\hline
N VI He$\delta$    & 23.277 & $23.175 \pm 0.004$ & \mbox{ $-1319 \pm 50$} & \mbox{ $637 \pm 85$} & \mbox{ $0.014 \pm 0.001$}  \\
N VI He$\gamma$   & 23.771 & $23.670 \pm 0.002$ & \mbox{ $-1268 \pm 37$} & \mbox{ $579 \pm 46$} & \mbox{ $0.026 \pm 0.002$}  \\
N VII Ly$\alpha$  & 24.779 & $24.706 \pm 0.010$ & \mbox{ $-1100 \pm 99$} & \mbox{ $739 \pm 116$} & \mbox{ $0.022 \pm 0.001$}  \\
N VI He$\beta$ r    & 24.898 & $24.806 \pm 0.004$ & \mbox{ $-1306 \pm 98$} & \mbox{ $638 \pm 73$} & \mbox{ $0.022 \pm 0.001$}  \\
C VI Ly$\epsilon$    & 26.026 & $25.922 \pm 0.003$ & \mbox{ $-1199 \pm 32$} & \mbox{ $650 \pm 58$} & \mbox{ $0.039 \pm 0.002$}  \\
C VI Ly$\delta$   & 26.357 & $26.245 \pm 0.003$ & \mbox{ $-1276 \pm 36$} & \mbox{ $617 \pm 66$} & \mbox{ $0.045 \pm 0.003$}  \\
C VI Ly$\gamma$  & 26.990 & $26.873 \pm 0.003$ & \mbox{ $-1294 \pm 35$} & \mbox{ $560 \pm 52$} & \mbox{ $0.068 \pm 0.005$}  \\
C VI Ly$\beta$  & 28.465 & $28.404 \pm 0.004$ & \mbox{ $-724 \pm 38$} & \mbox{ $1280 \pm 90$} & \mbox{ $0.157 \pm 0.009$}  \\
N VI He$\alpha$ r  & 28.787 & $28.660 \pm 0.003$ & \mbox{ $-1326 \pm 40$} & \mbox{ $700 \pm 55$} & \mbox{ $0.145 \pm 0.008$}  \\
Fe XXV?  & 29.4132$^c$   & $29.326 \pm 0.002$ & \mbox{ $-889 \pm 22$} & \mbox{ $439 \pm 41$} & \mbox{ $0.081 \pm 0.006$ } \\
\hline
\end{tabular}
\end{center}
Notes:\hspace{0.1cm} $^a $: Uncertain identifications are marked with question marks. $^b $: Time in days after the optical peak on August 16.25 UT, 2013 (Modified Julian Date (MJD) 56520.25). $^c $: The rest wavelength was obtained from the Chianti atomic database version 11.0 \citep{1997A&AS..125..149D, 2024ApJ...974...71D}. \\
\end{flushleft}
\end{minipage}
\end{table*}
\setcounter{table}{2}
\begin{table*}
\begin{minipage}{180mm}
\begin{flushleft}
\caption{Continued. Rest wavelength, observed wavelength, velocity, broadening velocity, and optical depth resulting from the fits of the absorption lines with proposed identification. The reported errors are at the 90\% confidence level. The rest wavelengths were obtained from AtomDB database version 3.0.9 \citep{2001ApJ...556L..91S, 2012ApJ...756..128F, 2020Atoms...8...49F}.}
\label{table:absorption V339}
\begin{center}
\begin{tabular}{lccccc}
\hline
Ion$^a$ & $\lambda_0$ & $\lambda_m$ & $v_{\rm shift}$ & $v_{\rm width}$ & $\tau_{\rm c}$  \\
& (\AA) & (\AA) & (km\,s$^{-1}$) & (km\,s$^{-1}$) &  \\
\hline
S XI   & 31.050 & $30.968 \pm 0.003$ & \mbox{ $-795 \pm 37$} & \mbox{ $913 \pm 49$} & \mbox{ $0.131 \pm 0.005$}  \\
Si XI  & 31.926 & $31.815 \pm 0.002$ & \mbox{ $-1046 \pm 30$} & \mbox{ $874 \pm 38$} & \mbox{ $0.230 \pm 0.005$}  \\
Fe XXIV & 32.377 & $32.271 \pm 0.004$ & \mbox{ $-982 \pm 42$} & \mbox{ $543 \pm 79$} & \mbox{ $0.094 \pm 0.013$}  \\
C V He$\delta$    & 32.754 & $32.609 \pm 0.005$ & \mbox{ $-1328 \pm 45$} & \mbox{ $615 \pm 82$} & \mbox{ $0.141 \pm 0.013$}  \\
C V He$\gamma$    & 33.426 & $33.295 \pm 0.002$ & \mbox{ $-1176 \pm 32$} & \mbox{ $671 \pm 30$} & \mbox{ $0.227 \pm 0.007$}  \\
C VI Ly$\alpha$  & 33.740 & $33.578 \pm 0.002$ & \mbox{ $-1441 \pm 39$} & \mbox{ $728 \pm 35$} & \mbox{ $0.264 \pm 0.009$}  \\
C V He$\beta$ r    & 34.973 & $34.811 \pm 0.003$ & \mbox{ $-1387 \pm 37$} & \mbox{ $732 \pm 44$} & \mbox{ $0.214 \pm 0.009$}  \\
Ar XI  & 36.244 & $36.098 \pm 0.002$ & \mbox{ $-1209 \pm 34$} & \mbox{ $483 \pm 31$} & \mbox{ $0.144 \pm 0.005$}  \\
S XI  & 39.240 & $39.089 \pm 0.006$ & \mbox{ $-1154 \pm 49$} & \mbox{ $392 \pm 89$} & \mbox{ $0.048 \pm 0.010$}  \\
C V He$\alpha$ r   & 40.267 & $40.079 \pm 0.004$ & \mbox{ $-1397 \pm 36$} & \mbox{ $756 \pm 72$} & \mbox{ $0.221 \pm 0.016$}  \\
S IX  & 47.433 & $47.234 \pm 0.009$ & \mbox{ $-1258 \pm 55$} & \mbox{ $799 \pm 107$} & \mbox{ $0.015 \pm 0.002$}  \\
S VIII?   & 52.756  & $52.568 \pm 0.017$ & \mbox{ $-1068 \pm 100$} & \mbox{ $864 \pm 135$} & \mbox{ $0.005 \pm 0.001$ } \\
Si IX  & 57.512 & $57.246 \pm 0.027$ & \mbox{ $-1384 \pm 145$} & \mbox{ $952 \pm 294$} & \mbox{ $0.003 \pm 0.001$}  \\
Fe XIV & 58.811 & $58.599 \pm 0.012$ & \mbox{ $-1080 \pm 63$} & \mbox{ $315 \pm 94$} & \mbox{ $0.002 \pm 0.001$}  \\
S VIII & 59.882 & $59.666 \pm 0.014$ & \mbox{ $-1083 \pm 72$} & \mbox{ $515 \pm 144$} & \mbox{ $0.002 \pm 0.001$}  \\
\hline
\end{tabular}
\end{center}
Notes:\hspace{0.1cm} $^a $: Uncertain identifications are marked with question marks. $^b $: Time in days after the optical peak on August 16.25 UT, 2013 (Modified Julian Date (MJD) 56520.25). $^c $: The rest wavelength was obtained from the Chianti atomic database version 11.0 \citep{1997A&AS..125..149D, 2024ApJ...974...71D}. \\
\end{flushleft}
\end{minipage}
\end{table*}

\section{Discussion}
\label{sec:discussion}

The underlying cause of the short-period (10 $-$ 100 s) oscillations observed in novae and persistent SSSs may be linked to WD rotation. However, this interpretation struggles to account for the transient nature of the signal, the slight variation in period, and the fact that such oscillations are detected in only a small fraction of systems \citep{2015A&A...578A..39N}. Since the rotation period of the white dwarf cannot change with the observed patterns, it is challenging for models to explain the variations in observed period. An intriguing explanation was used to interpret these oscillations: nonradial g-mode pulsations of the WD driven by the $\epsilon$ mechanism \citep{2003ApJ...584..448D, 2006AJ....131..600S, Osborne2011, 2015A&A...578A..39N}. However, \citet{2018ApJ...855..127W} noted that explaining the 54-second period in V339 Del through g-modes is challenging, as the unstable surface g-modes typically have shorter periods than those observed in novae.

The modulation amplitude and duration of the short-term periodicity observed in the X-ray light curve of V339 Del exhibit temporal variations. \citet{Orio2021} suggested that the amplitude and period of the X-ray light curve variations in nova LMC 2009a could sometimes make the short-term period undetectable, though they are likely always present. However, this may not apply to V339 Del. \citet{2015A&A...578A..39N} proposed that short-term periodicity in some super-soft novae and persistent super-soft sources might be a transient phenomenon, which appears to be the case for V339 Del. This nova shows a more pronounced period signal, a larger fraction of total observed time with significant detections, and a greater amplitude compared to nova LMC 2009a. \citet{2015A&A...578A..39N} considered whether Thomson scattering could degrade the amplitude (and thus detectability) of periodic signals, which might explain the rarity and transient nature of such signals. They found that Thomson scattering does not significantly impact the detectability of periodic signals. We suggest the interpretation that the observed variations in amplitude and the transient nature of the short-term periodicity in V339 Del may be due to temporary obscuration events affecting the emission from the central hot source.

The X-ray spectra of V339 Del exhibit a blackbody-like continuum with clearly defined absorption lines and lack distinct emission lines, categorizing them as SSa (a for absorption lines) spectra based on the SSa subclass definition by \citet{2013A&A...559A..50N}. Similar SSa spectra are found in persistent SSS sources such as Cal 83 and RX J0513.9-6951, classical novae like LMC 2012, V4743 Sgr, KT Eridani, and V2491 Cyg, as well as recurrent nova RS Oph \citep{2013A&A...559A..50N}. Atmosphere models could be fit to the SSa spectra or have been used to fit the SSa spectra with some success, and it is possible that all SSa spectra can be explained by these models \citep{2013A&A...559A..50N}. In contrast, the X-ray spectra of SSS Cal 87 \citep{2024A&A...690A...9P} and recurrent nova U Scorpii \citep{2012ApJ...745...43N} differ significantly from V339 Del, featuring a weak blackbody-like continuum without absorption lines and with emission lines that are at least as strong as the continuum, especially where the continuum is most intense. These spectra are classified as the SSe (e for emission lines) subclass by \citet{2013A&A...559A..50N}.

\citet{2013A&A...559A..50N} suggested that the emission lines observed in SSS Cal 87 and recurrent nova (RN) U Scorpii might also be present in SSa spectra, hidden within a complex array of atmospheric absorption and emission features. A similar scenario may apply to V339 Del's spectra, which are classified as SSa spectra. It is possible that some emission lines are present but difficult to discern within the complex atmospheric spectrum of V339 Del; however, no concrete evidence has been found to support this at present. For most novae and SSS, emission lines are typically attributed to the ejecta \citep{2013A&A...559A..50N}. However, in contrast to the X-ray spectra of KT Eridani, which exhibit emission lines, V339 Del's spectra show no emission lines, indicating that V339 Del's spectrum is much closer to being predominantly dominated by the atmosphere component, much less stuff from the ejecta.

\section{Summary and Conclusions}
\label{sec:conclusions}
The spectra of novae in the X-ray supersoft phase are intricate but provide a wealth of physical information. Our analysis reveals that the spectra of V339 Del are primarily dominated by atmospheric continuum emission with clearly visible absorption lines, and no emission lines are present. The X-ray spectra of V339 Del are SSa spectra. Interstellar absorption lines of O I, N I, and C II are identified in all three observations, appearing at their rest wavelengths. Other identified absorption lines exhibit blue shifts, with velocities ranging from $\sim$ 724 to $\sim$ 1474 km s$^{-1}$ and broadening velocities ranging from 310 to 1280 km s$^{-1}$; the majority of these lines have velocities approximately around 1200 km s$^{-1}$. The highly structured nature of SSa spectra, which no current model can fully reproduce \citep{2013A&A...559A..50N}, also applies to the X-ray spectra of V339 Del. Future fitting models should account for absorption features arising from layers with varying optical depths and velocities. We calculated the velocity of the strongest absorption features and the optical depths at which they occur, providing valuable data for developing more adequate physical models in the near future.

Although the spectra of V339 Del do not exhibit large irregular variability of prominent emission lines like those seen in KT Eridani, the analysis is complicated by light curve and spectral variations on the order of a few thousand seconds. Additionally, potential emission lines may be obscured within the complex atmospheric continuum emission and absorption features.

Finally, we confirmed that the SSS pulsations with a $\sim$ 54 s timescale were detected on 97.0 and 112.0 days after the optical maximum, when the X-ray count rate varied by a factor of 1.25 to 1.31. This behaviour contrasts with RS Oph, where a $\sim$ 35 s periodicity was observed only during the first $\sim$55 days after optical maximum in the 2006 outburst \citep[see][]{Nelson2008, Osborne2011, 2013A&A...559A..50N} and throughout the entire SSS phase in the 2021 outburst \citep{2023ApJ...955...37O}. The $\sim$ 54 s pulsation in V339 Del exhibited a transient nature, with periodic signals being strong when present but varying in amplitude over timescales of a few thousand seconds. This characteristic suggests that any physical explanation for this phenomenon, also observed in other novae, must account for these variations. We propose that temporary obscuration events affecting the emission from the central hot source may be responsible. We also confirmed the drift of this $\sim$ 54 s period. The significant variations in the pulse profiles, along with the fact that the phase-folded light curve for the full observations on days 97.0 and 112.0 exhibits notable departures from a pure sine wave, may be associated with this period drift. There is weak evidence for an anti-correlation between the $\sim$ 54 s period modulation amplitudes and the count rates on day 97.0. This anti-correlation may hint at temporary obscuration events that intermittently block the central source. If the pulsations originate near the WD surface, changes in obscuration could dampen the periodic signal more significantly than the overall count rate, as pulsations could be more sensitive to local conditions. This interpretation aligns with the complex and evolving nature of post-nova environments and could serve as a basis for future, more detailed investigations.

\section*{Data Availability}
The data analyzed in this article are all available in the HEASARC archive of NASA at the following URL: \url{https://heasarc.gsfc.nasa.gov/db-perl/W3Browse/w3browse.pl}

\section*{Acknowledgements}
We extend our sincere gratitude to reviewers for the insightful and constructive comments, which helped us to improve the scientific content of this article. We acknowledge with thanks the observers worldwide for contributing their data to the AAVSO. This work has made use of data obtained with the gratings on board {\it Chandra} and {\it XMM-Newton}. This work also made use of data supplied by the UK {\it Swift} Science Data Centre at the University of Leicester. Songpeng Pei thanks Marina Orio for many useful conversations. Song-Peng Pei is supported by the High-level Talents Research Start-up Fund Project of Liupanshui Normal University (LPSSYKYJJ202208), the Science and Technology Foundation of Guizhou Province (QKHJC-ZK[2023]442) and the Discipline-Team of Liupanshui Normal University (LPSSY2023XKTD11). Nataly Ospina acknowledge funding from the Ministry of Universities of Spain, the Recovery, Transformation and Resilience Plan (PRTR), and the UAM though the Grant CA3/RSUE/2021-00559 and the Grant PID2021-124050NB-C31 funded by MCIN/AEI/10.13039/501100011033 and by the European Union NextGenerationEU/PRTR and the European Union’s Horizon 2020 Research and Innovation Programme under the Marie Sklodowska-Curie 2020-MSCA-RISE-2019 SK2HK grant agreement no. 872549. Qiang Li is supported by the Research Foundation of Qiannan Normal University for Nationalities (No.QNSY2019RC02). Zi-wei Ou is supported by the National Natural Science Foundation of China (NSFC, Grant No. 12393853). Tao-Zhi Yang is supported by the program of the National Natural Science Foundation of China (grant Nos. 12003020) and Shaanxi Fundamental Science Research Project for Mathematics and Physics (Grant No. 23JSY015). Yong-Zhi Cai is supported by the National Natural Science Foundation of China (NSFC, Grant No. 12303054), the Yunnan Fundamental Research Projects (Grant No. 202401AU070063) and the International Centre of Supernovae, Yunnan Key Laboratory (No. 202302AN360001).




\bibliographystyle{mnras}
\bibliography{V339Del} 




\bsp	
\label{lastpage}
\end{document}